\documentclass[aps,graphics,a4paper,10pt,twocolumn]{revtex4-2}

\usepackage[margin=25mm]{geometry}

\usepackage{amsmath}
\usepackage{braket}
\usepackage{hyperref}
\usepackage{textcomp}
\usepackage{xcolor}

\usepackage{tikz}
\usepackage{pgfplots}
\usepackage{tikzscale}
\usepackage{bbm}
\usepackage{commath}

\usepackage{soul}

\usepackage[title,page]{appendix}

\newcommand{\be}{\begin{equation}\begin{aligned}} 
\newcommand{\ee}{\end{aligned}\end{equation}}
\newcommand{\beN}{\begin{equation*}\begin{aligned}} 
\newcommand{\eeN}{\end{aligned}\end{equation*}}
\newcommand{\ba}{\begin{array}}
\newcommand{\ea}{\end{array}}
\newcommand{\bqa}{\begin{eqnarray}}
\newcommand{\eqa}{\end{eqnarray}}
\newcommand{\RN}[1]{
  \textup{\expandafter{(\romannumeral#1)}}
}
\newcommand{\eq}[1]{Eq.~\eqref{eq:#1}}

\newcommand{\na}{\frac{N}{2}}

\newcommand{\bbrkts}[1]{\Big( #1 \Big)}
\newcommand{\brkts}[1]{\left( #1 \right)}

\newcommand{\fiveT}{{\cal T}}
\newcommand{\ft}{{\cal T}}

\newcommand{\new}[1]{{\color[rgb]{0, 0, 0}{#1}}}

\begin{document}

\title{Deterministic preparation of non-classical states of light in cavity-optomechanics}

\author{Yuxun Ling}
\author{Florian Mintert}

\affiliation{Physics Department,	Blackett Laboratory, Imperial College London, Prince Consort Road, SW7 2BW, United Kingdom}
\date{\today}

\begin{abstract}
Cavity-optomechanics is an ideal platform for the generation non-Gaussian quantum states due to the anharmonic interaction between the light field and the mechanical oscillator;
but exactly this interaction also impedes the preparation in pure states of the light field.
In this paper we derive a driving protocol that helps to exploit the anharmonic interaction for state preparation, and that ensures that the state of the light field remains close-to-pure.
This shall enable the deterministic preparation of photon Fock states or coherent superpositions thereof.
\end{abstract}

\maketitle

\section{Introduction}

Optomechanical experiments provide accurate control over the quantum dynamics of mesoscopic mechanical oscillators and light fields on the single photon level~\cite{AKM14}.
In particular because the interaction between such oscillators and light fields is anharmonic,
there is great potential to generate non-classical, non-Gaussian states~\cite{KDM+10, VPC+11, Rab11, BBS+12, LM18,XSL+19}
with various applications including quantum metrology~\cite{HWH+14,HWD08, DJK15}, quantum cryptography~\cite{GRT+02, YLC+10, ACF+09}, and more~\cite{MLG+06,NFC09, DSA+10,ZSS18}.
A widely pursued goal is the creation of single photon Fock states \cite{Rab11,LDC16,SS18, SZX+18,WBL+19, WBH+20},
but also multi-photon Fock states and superposition of Fock states
are of use in practical applications~\cite{DSB02, NOO+07, LSP+02, OJT+07, DDI06}. 

While the anharmonic interaction is essential for the realization of non-Gaussian states,
its flip side is that it results in the evolution towards quantum states with correlations between the optical and mechanical degrees of freedom.
The preparation of a pure state of one of the subsystems can be achieved in terms of a projective measurement on the other subsystem~\cite{KDM+10, VPC+11, MFB19}.
Such a scheme, however, is intrinsically probabilistic with a success probability limited by the multitude of possible measurement outcomes.
In this work, we propose a driving scheme for optomechanical systems for the deterministic preparation of close-to-pure, non-classical states of light,
and we exemplify the scheme with two-photon Fock states and the coherent superposition of this state and the vacuum state.

The main body of the paper is divided into two sections. 
In Sec.\ref{sec:theory}, a perturbative solution to the driven evolution of the system is derived and the optimal driving protocol is constructed. 
In Sec.\ref{sec:examples}, numerical results for both the coherent dynamics and the dissipative dynamics are presented and analysed.

\section{Theory}\label{sec:theory}

\subsection{System Hamiltonian}

We consider a generic model of an opto-mechanical system composed of a Fabry–P\'erot cavity with one stationary and one movable mirror, {\it i.e.} a mechanical oscillator. 
The system Hamiltonian $H=H_{0}+H_{I}$ consists of the non-interacting part $H_{0}=\omega_c a^\dag a+\omega_m b^\dag b$ with resonance frequencies $\omega_c$ and $\omega_m$ of the optical field and the mechanical oscillator, as well as corresponding creation and annihilation operators a, b, $a^\dagger$ and $b^\dagger$. 
The interaction $H_{I}=-g_0 a^\dag a(b^\dag+b)$ is cubic in these operators, and thus can overcome the restriction to Gaussian dynamics that is inherent to quadratic Hamiltonians.

Pumping of the cavity with an external classical light field is described by the Hamiltonian
\be\label{eq:drivHam}
H_{d}=d(t)a^\dagger+d^*(t)a\ ,
\ee
where the time-dependent function $d(t)$ characterizes the frequency and any other time-dependence of the light field, such as a temporal modulation.

\new{In the absence of the interaction $H_I$, driving the cavity with classical light can only result in the creation of a classical state. 
It is only the interaction that can result in nonclassical features, as indicated for example by negative values of the Wigner function~\cite{KZ04}.
In fact, the interacting system is equivalent to an anharmonic oscillator with a quartic Kerr nonlinearity which can be used for the generation of a variety of non-Gaussian states~\cite{Leo96,RGR13}.
This nonlinearity, however, is not a single-body property of the light field, but it is a property of a collective degree of freedom which is defined by the polaron transformation $e^{(g_0/\omega_m)a^\dag a(b^\dag-b)}$.
This highlights that the optomechanical interaction will generally result in correlations between the optical and mechanical degrees of freedom.
Each of these parts of the system alone is then described by a mixed states, and this mixing typically prevents the observation of nonclassical effects~\cite{GZ20}.}

A way around this predicament is to perform a measurement on the light field resulting in the projection of the mechanical oscillator into a pure quantum state.
Such a probabilistic method is, however, not easily generalizable to the creation of states of light, because the mechanical degree of freedom is substantially less accessible for a measurement.

For the creation of pure, non-classical states of light it is thus highly desirable to ensure that the state of the full system is given as direct product of the states of each of the subsystems.
It is possible to deterministically create highly non-classical states of the mechanical oscillator in terms of suitably shaped pump profiles \cite{LM18}, but the extension of such an approach for the creation of states of light is rather challenging, since the control of the system is realized in terms of driving
the light field, which needs to interact with the mechanical oscillator before excitations in the mechanical oscillator can cause the light field to adopt non-classical features.
At the same time, the driving tends to create Gaussian features of the state of light that tend to over-shadow non-classical effects.

\subsection{Driving profiles}

The range of achievable states will strongly depend on the chosen driving profile $d(t)$.
In particular strong driving can modify the system dynamics with great potential for state preparation, but practical constraints demand sufficiently weak driving with sufficiently simple spectra.

Suitable choices for driving profiles can be identified from the basic properties of the opto-mechanical interaction $a^\dag a(b^\dag+b)$
that couples the absorption and emission of a phonon to the photon number operator $n_c=a^\dag a$.
Since the number operator is quadratic, and thus does not contribute to the creation of non-classicality,
it will not be helpful to use driving that supports the absorption and emission of single phonons.
An effective process, comprised of two successive interaction events, on the other hand would involve the operator $n_c^2$, which is no longer quadratic.
We will thus employ a driving profile that favours the absorption and emission of pairs of phonons.

A reasonably elementary driving profile satisfying this requirement is given by
\be
d(t)=2\new{i}Ee^{-i(\omega_ct-\psi)}\cos(2\omega_mt)\ ,
\label{eq:drivingprofile}
\ee
with a detuning from the optical resonance frequency $\omega_c$ that amounts to twice the mechanical frequency $\omega_m$.
The real-valued amplitude $E$ and phase $\psi$ are not determined yet, and the freedom to choose these parameters will be utilized for the design  of suitable driving patterns.

\subsection{System dynamics}
Given the cubic character of the interaction, it is not possible to solve the dynamics of the driven-interacting system exactly.
Since, however, the non-interacting system is described by a quadratic Hamiltonian, the dynamics of the driven, but non-interacting system can be found analytically.
It is thus natural to solve the system dynamics perturbatively in the interaction strength.

The propagator of the non-interacting system is given by~\cite{MMT97}
\be\label{eq:U0}
U_0(t)&=e^{i\xi(t)} e^{-i\omega_cn_ct}e^{-i\omega_mn_mt}e^{f(t)a^\dag-f^*(t)a}\ ,\\
\ee
with \new{the phonon number operator $n_m=b^\dag b$,
the time-dependent scalar function}
\be
f(t)&=\int_{0}^{t}d\tau\ e^{i\omega_c\tau}d(\tau) \new{\ ,}
\ee
and a real, scalar, time-dependent global phase $\xi(t)$.

The interaction Hamiltonian $\tilde{H}_{I}=U_0^\dag H_{I}U_0$ in the frame defined by $U_0$ reads
\be\label{eq:HItilde}
\tilde{H}_{I}=
-g_0\bbrkts{a^\dag+f^*(t)}\bbrkts{a+f(t)}
\bbrkts{b(t)+b^\dag (t)}\new{\ ,}
\ee
with the time-dependent annihilation operator $b(t)=be^{-i\omega_mt}$ and creation operator $b^\dag(t)=b^\dag e^{i\omega_mt}$ of a phonon.

Since the cavity frequency $\omega_c$ is much larger than the frequency $\omega_m$ of the mechanical oscillator, one can take $\omega_c$ to be an integer multiple of $\omega_m$.
In this case, the propagator of non-interacting system $U_0$ is periodic with period $T=2\pi/\omega_m$.
Furthermore, the Hamiltonian $\tilde{H}_{I}$ also becomes periodic, and the propagator induced by $\tilde{H}_{I}$ thus admits a decomposition into a periodic part that reduces to the identity $\mathbbm{1}$
after full periods and a part $\exp(-iH_et)$ that is induced by the effective Hamiltonian $H_e$ which fully captures the dynamics of the system after full periods.

The effective Hamiltonian for the evolution after one mechanical period can be constructed perturbatively~\cite{Mag54} in powers of $k=g_0/\omega_m$.
It will be insightful to distinguish between the parts of the effective Hamiltonian that capture processes of only the cavity field, processes of only the mechanical oscillator and interaction processes each.
That is, the explicit expansion of the effective Hamiltonian is given by
\be\label{eq:mag2nd}
H_e=\frac{\omega_m}{2}\sum_jk^j\left(\mathcal{M}_j^C+\mathcal{M}_j^M+\mathcal{M}_j^I\right)\ ,
\ee
with the symbols $C$, $M$ and $I$ referring to cavity, mechanical and interaction.

With the explicit driving profile given in \eq{drivingprofile} the lowest order effective Hamiltonian (linear in $k$) vanishes exactly,
and the second order contribution 
is the dominant term.

Also the second order term $\mathcal{M}_2^M$ of the mechanical oscillator vanishes,
but the term $\mathcal{M}_2^C$ is generally finite and reads
\be\label{eq:cavity2nd}
\mathcal{M}_2^C=-2n_c^2-\frac{4\eta^2}{3}n_c+\frac{\eta^2}{3}(a^2e^{-2i\psi}+\mbox{h.c.})\ ,
\ee
with the scaled driving amplitude $\eta=E/\omega_m$. 
The second order interaction term reads
\be\label{eq:interaction2nd}
\mathcal{M}_2^I=-\sqrt{2}\eta P_\psi\bbrkts{b^2+(b^\dag)^2}\ ,
\ee
with the phase-shifted momentum
\be\label{eq:cavitymomentum}
P_\psi=\frac{i}{\sqrt{2}}(ae^{-i\psi}-a^\dag e^{i\psi})
\ee
of the cavity field.

\subsection{Effective Hamiltonian}\label{sec:EffHam}

The cavity operator $\mathcal{M}_2^C$ in \eq{cavity2nd} contains several terms that are very useful for the generation of non-Gaussian states.
The terms $(a^\dag)^2$ and $a^2$ describe the creation and annihilation of pairs of photons.
Those processes alone, however, are still within the set of Gaussian dynamics, but the quartic operator $n_c^2$ breaks this restriction.
It results in a deviation from the evenly spaced level structure of the quantum harmonic oscillator.
The effectively larger spacing between higher lying levels makes it possible to populate the two-photon Fock state starting from the vacuum state, while making sure that population of the four-photon Fock state and higher-lying states are sufficiently far off-resonant to be negligible.

On the other hand, in second order there are also processes that can impede the generation of non-Gaussian states.
The term linear in $n_c$ that is contained in the cavity operator $\mathcal{M}_2^C$ increases the constant spacing between energy levels of the quantum harmonic oscillator and thus
reduces the effective anharmonicity resulting from the $n_c^2$ term. 
Furthermore, the interaction term $\mathcal{M}_2^I$ in \eq{interaction2nd} results in correlations between the cavity and the mechanical oscillator to build up.

With the driving profile specified in \eq{drivingprofile}, it is not possible to have these undesired terms vanish without also having terms describing the creation and annihilation of pairs of photons vanish.
Goal of the following analysis will therefore be the construction of a sequence of driven and undriven intervals such that the
effective Hamiltonian \new{$H_f$} for the dynamics over all these intervals
contains the desired terms, but in which terms describing undesired processes are no longer present.

\new{The dynamics of each period of driven dynamics is characterized in terms of an effective Hamiltonian $H_e$ given in Eq.~\eqref{eq:mag2nd},
but since the driving can vary from period to period, there is a distinct effective Hamiltonian for each period.
Those effective Hamiltonians
can then be taken as starting point for the construction of an effective Hamiltonian $H_f$
that characterizes the dynamics of several individual periods of driven dynamics.}
Because of non-commutativity, the effective Hamiltonian $H_f$ would need to be constructed with the Baker-Campbell-Hausdorff series.
However, because the leading contribution to the individual effective Hamiltonian is of order $k^2$, the first non-trivial contribution to the Baker-Campbell-Hausdorff series is of order $k^4$, which is smaller than the highest order ({\it i.e.} $k^2$) that is included in the individual effective Hamiltonians so far, and smaller than the highest order (\new{\it i.e.} $k^3$) that will be included later-on.
Within the given level of accuracy, the complete effective Hamiltonian $H_f$ is thus given by the sum of the individual effective Hamiltonians.

The interaction term $\mathcal{M}_2^I$ (\eq{interaction2nd}) depends on the phase $\psi$ of the driving field (\eq{drivingprofile}) via the momentum operator $P_\psi$ given in \eq{cavitymomentum}.
After $N$ periods of driven dynamics, with driving strength $\eta_j$ and phase $\psi_j$ in period $j$, the interaction term in the full effective Hamiltonian thus contains the factor $\sum_{j=1}^N\eta_jP_{\psi_j}$.
Any series of driving periods satisfying $\sum_{j=1}^{N}\eta_j\exp\{i\psi_j\}=0$
can thus ensure that there are no interaction effects in leading order at the end of the dynamics.

The creation and annihilation of pairs of photons in \eq{cavity2nd}, on the other hand depend on the
driving field via $\eta^2\exp\{\pm 2i\psi\}$, and one can easily find choices for the driving fields
such that $\sum_j^{N}\eta_j^2\exp\{2i\psi_j\}$ is finite while the condition $\sum_j^{N}\eta_j\exp\{i\psi_j\}=0$ is satisfied.
To leading order, this prescription would result in an effective Hamiltonian

\be
\label{eq:effHamForPsi}
H_{f}=\frac{\omega_m}{2}
			 Nk^2\left(
					\left(\frac{\zeta}{3} a^2+\mbox{h.c.}\right)-2n_c^2-\frac{4}{3}\chi n_c
			\right)\ ,
\ee
with
\be
\zeta=\frac{1}{N}\sum_{j=1}^{N}\eta_j^2 e^{-2i\psi_j}\ ,\ \mbox{and}\
\chi=\frac{1}{N}\sum_{j=1}^{N}\eta_j^2\ .
\ee

This is a viable effective Hamiltonian for the creation of non-classical states, with a non-linearity that breaks the restriction to Gaussian dynamics.
Yet, in practice, it is desirable to have a non-linearity $n_c^2$ that is strong as compared to the linear term $\propto n_c$ in order to obtain a spectrum of the diagonal part of $H_f$ in \eq{effHamForPsi} with strongly un-even spacing between neighbouring energy levels.
Since the linear part $\propto n_c$ in \eq{effHamForPsi} is getting strong as compared to the non-linear part $\propto n_c^2$ in the regime of strong driving, it is necessary to find a mechanism that effectively reduces this linear part.

The central idea that allows to achieve this, is that the free evolution induced by $n_c$ results in the type of phase shift described by the $\psi_j$.
That is, a ramp in the phases $\psi_j$ has approximately the same effect as a true free phase evolution.

In order to formalize this,
it is helpful to note that the effective Hamiltonians $\mathcal{M}_2^{C}(\psi)$ and $\mathcal{M}_2^{I}(\psi)$ in Eqs.\eqref{eq:cavity2nd} and \eqref{eq:interaction2nd} satisfy the relation
\be
\mathcal{M}_2^{C/I}(\psi)=V_\psi\mathcal{M}_2^{C/I}(0)V_\psi^\dagger\ ,
\ee
with $V_\psi=e^{i\psi n_c}$.
The propagator induced by the effective Hamiltonian can thus be written as
\be\label{eq:phaseTrans}
U_\psi=e^{-iH_e(\psi)T}=V_\varphi e^{-iH_e(\psi-\varphi)T}V_\varphi^\dagger\ ,
\ee
with a phase $\varphi$ that can be chosen at will.

Because of the identity $V_{\varphi_{j+1}}^\dagger V_{\varphi_{j}}=V_{\phi_j}^\dagger$,
with $\phi_j=\varphi_{j+1}-\varphi_{j}$,
the product of two propagators of consecutive periods simplifies to
\bqa\label{eq:productOfTwoPeriods}
U^{(j+1)}U^{(j)}&=&
V_{\varphi_{j+1}} e^{-iH_e(\psi_{j+1}-\varphi_{j+1})T}V_{\phi_j}^\dagger\times\nonumber\\
&&\times e^{-iH_e(\psi_j-\varphi_{j})T}V_{\varphi_{j}}^\dagger\ .
\eqa
The full propagator over $N$ periods can thus be expressed as
\bqa
U&=&V_{\varphi_{N+1}}\prod_{j=1}^{N}V_{\phi_j}^\dagger e^{-iH_e(\psi_j-\varphi_{j})T}\ .
\label{eq:product}
\eqa

The first factor $V_{\varphi_{N+1}}$ describes a free phase evolution after all the time-intervals of driven dynamics.
Since this is merely a rotation in phase space, it has no bearing on the classical or quantum mechanical character of the final quantum states.

Each factor $V_{\phi_j}^\dagger e^{-iH_e(\psi_j-\varphi_{j})T}$ in \eq{product} is a product of a term $e^{-iH_e(\psi_j-\varphi_{j})T}$ induced by the effective Hamiltonian and a term of free phase evolution induced by $n_c$.
In the limit of infinitesimally short intervals $T\to 0$, this is equivalent to an evolution induced by an effective Hamiltonian with a modified term $n_c$.
In practice, the duration $T$ will always be finite, but the approximation
\be
V_{\phi_j}^\dagger e^{-iH_e(\psi_j-\varphi_{j})T}\simeq e^{-i(H_e(\psi_j-\varphi_{j})T\new{)}+i\phi_j n_c}
\ee
is sufficiently good for the purpose of state preparation for realistic values of $T$.

Now the effective Hamiltonian for each period contains an extra term $\propto n_c$ in addition to $H_e$,
and the freedom to choose values for the phase angles $\psi_j$ and $\phi_j$ can be used to ensure that undesired terms cancel.
For any choice of the driving parameters $\eta_j$ and $\psi_j$ satisfying
\be
\label{eq:finPhaseCond}
\sum_{j=1}^{N}\eta_je^{-i(\psi_j-\varphi_{j})}=0\ ,\\
\sum_{j=1}^{N}\eta_j^2e^{-2i(\psi_j-\varphi_{j})}\neq0\ , \\
\ee
the effective Hamiltonian after $N$ periods will reduce to
\be\label{eq:HeffSO+phaseAngles}
&\frac{\omega_m}{2}k^2 N\left(\left(\frac{\zeta'}{3} a^2+\mbox{h.c.}\right)-2n_c^2\right)-\\
&-\frac{\omega_m}{2}\sum_{j=1}^N\left(\left(\frac{4}{3}k^2\eta_j^2-\frac{\phi_j}{\pi}\right)n_c\right)+O(k^2\phi) \ ,
\ee
with 
\be
\zeta'=\frac{1}{N}\sum_{j=1}^{N}\eta_j^2e^{-2i(\psi_j-\varphi_{j})} \new{\ .}
\ee
With the specific choice of $\phi_j=4/3\ \pi k^2\eta_j^2$, terms linear in $n_c$ will vanish in the leading order during each period. 
Since $\phi_j$ is the accumulated phase shift of the driving pattern,
this amounts to the individual phases
\be
\varphi_j=\frac{4}{3}\pi k^2\sum_{l=1}^{j-1}\eta_l^2
\label{eq:phase}
\ee
that increase by an amount determined by the driving amplitude $\eta_j$.

\subsection{Driving pattern}
\label{sec:drivingpattern}

While, in principle, it is only required that the effective Hamiltonian for the dynamics over the entire interval of interest matches the desired Hamiltonian,
it is preferable that such a condition is satisfied at in-between points in time.
We will therefore consider driving protocol in which the full time-window of duration $NT$ is divided into $\na$ blocks of duration $2T$,
and require that the dynamics over each block is induced by the desired effective Hamiltonian within the perturbative approximation.

In each such block, we will consider a constant driving amplitude, so that
the phases
\be\label{eq:phasejumps}
\psi_{2j}&=\pi+\frac{8}{3}\pi k^2\sum_{l=1}^{j-1}\eta_{2l}^2\ , \\
\psi_{2j+1}&=\frac{4}{3}\pi k^2\brkts{\eta_{2j}^2+2\sum_{l=1}^{j-1}\eta_{2l}^2}
\ee
of the driving profiles are suitable solutions of \eq{phase}.
With this choice the resulting effective Hamiltonian over $N$ periods of driven dynamics reads
\be
\label{eq:effHam2ndFin}
H_{g}=\omega_mk^2\sum_{j=1}^{\frac{N}{2}}\left(\frac{2}{3}\eta_{2j}^2\left(a^2+(a^\dag)^2\right)-4n_c^2\right)\ .
\ee
This can be taken as a starting point for state preparation, but it is worth exploring higher order perturbative corrections, since this will help to substantially increase the accuracy of state preparation with only slightly more involved driving patterns.

\subsection{Third Order Corrections}

The third order term $\mathcal{M}_3^C$ of the cavity vanishes,
and the term $\mathcal{M}_3^M$ of the mechanical oscillator reads
\be
\mathcal{M}_3^M=y\eta^2\left(b^\dag\cos\frac{5\pi}{12} +b \sin\frac{5\pi}{12}\right)^3+\mbox{h.c.}\ ,\\
\ee
with $y=16\sqrt{6}/27\simeq 1.45$.

The third order interaction term $\mathcal{M}_3^I$ is of the form
\be
\mathcal{M}_3^I=A_\psi X_m+B_\psi P_m+X_\psi G_m\ ,
\label{eq:Heffthirdorder}
\ee
with
\be
A_\psi&=\frac{\sqrt{2}}{15}\bbrkts{36\eta^3P_\psi+35\eta(n_cP_\psi+P_\psi n_c)}\ ,\\
B_\psi&=2\sqrt{2}i\eta^2\bbrkts{(a^\dag)^2 e^{2i\psi}-a^2e^{-2i\psi}}\ ,\\
G_m&=\frac{3yi}{4}\eta\left(b^\dag\cos\frac{5\pi}{12} -b \sin\frac{5\pi}{12}\right)^3+\mbox{h.c.}\ ,\\
\ee
and
\be
X_\psi&=\frac{1}{\sqrt{2}}(ae^{-i\psi}+a^\dag e^{i\psi})\ ,\\
X_m&=\frac{1}{\sqrt{2}}(b+b^\dag)\ , \\
P_m&=\frac{i}{\sqrt{2}}(b-b^\dag)\ .
\ee

With the driving pattern derived above in Sec.\ref{sec:drivingpattern}, the effective Hamiltonian for the dynamics over two periods of driving with constant driving amplitude is given by
\be\label{eq:thirdOrderHamWithCorrPhase}
H_p=H_g+\omega_m k^3(\mathcal{M}_3^M+B_0 P_m) \ ,
\ee
in which $H_g$ is the second order effective Hamiltonian in \eq{effHam2ndFin} and $B_0$ is the operator $B_\psi$ evaluated at $\psi=0$.
The contributions from $A_\psi$ and $X_\psi$ in $\mathcal{M}_3^I$ (\eq{Heffthirdorder}) average out in the effective Hamiltonian $H_p$.
However, there is still a finite correction due to the term $B_0 P_m$ in the interaction $\mathcal{M}_3^I$;
and a contribution to the effective Hamiltonian from the term $\mathcal{M}_3^M$ of the mechanical oscillator.

In addition to the interaction term $B_0P_m$, it is also desirable to remove other term $\mathcal{M}_3^M$ from the full effective Hamiltonian because it induces excitations in the mechanical oscillator, and the impact of interactions on the optical field remaining in higher order corrections tends to be increasing with growing excitations of the mechanical oscillator.
Therefore, we devote the following paragraphs to modifying the driving pattern such that these terms are reduced as much as possible.

The basic idea is that half a period of free evolution of the mechanical oscillator corresponds to a phase-shift of $\pi$ in the creation and annihilation operators $b^\dagger$ and b.
Since $\mathcal{M}_3^M+B_0P_m$ is an odd function in b and $b^\dagger$, a sequence of two periods of dynamics induces by $\mathcal{M}_3^M+B_0P_m$ with an in-between half-period of free evolution results in an effective cancellation.
Any driving protocol with alternating driven dynamics and intervals of free evolution,
with the phase of the driving fields satisfying \eq{phasejumps},
and the duration of free evolution being half a period of free evolution
 thus realises the dynamics described in Sec.\ref{sec:drivingpattern}, and it ensures cancellation of the dominant terms deviating from the desired effective Hamiltonian.

In practice, however, it is not possible to realize an exact free evolution of the mechanical oscillator because of the intrinsic interaction between the mechanical oscillator and the light field.
As we will see in the following, an interval of un-driven dynamics can be used to achieve a similar effect.

Following \eq{U0} the propagator for the undriven dynamics over half a mechanical period reads
\be\label{eq:UHalfInterv}
U_{\frac{T}{2}}=e^{\pi ik^2n_c^2}e^{-2\sqrt{2}ikn_cP_m}e^{-\pi in_m}\ ,
\ee
given that the optical resonance frequency is an even multiple of mechanical resonance frequency. 
The adjoint of this propagator satisfies the relation
\be
U_{\frac{T}{2}}^\dagger=U_{\frac{T}{2}}e^{-2\pi ik^2n_c^2}\ .
\ee

Any sequence including two intervals of driven dynamics and an interval of un-driven dynamics before each interval of undriven dynamics thus results in the propagator
\be
U_dU_{\frac{T}{2}}U_dU_{\frac{T}{2}}&=U_dU_{\frac{T}{2}}^\dagger e^{2\pi ik^2n_c^2} U_dU_{\frac{T}{2}}\\
&\simeq e^{-iH_pT}U_{\frac{T}{2}}^\dagger e^{2\pi ik^2n_c^2-iH_pT}U_{\frac{T}{2}}\ ,
\ee
where terms of order $k^4$ in the Baker-Campbell-Hausdorff relation are neglected.
Within the same approximation, the propagator over the four intervals reads

\be
\simeq \exp\brkts{-iT\brkts{H_p+U_{\frac{T}{2}}^\dagger H_pU_{\frac{T}{2}} }+2\pi ik^2 n_c^2}\ .
\ee

To leading orders ({\it i.e.} $k^3$), the term $U_{\frac{T}{2}}^\dagger (\mathcal{M}_3^M+B_0 P_m)U_{\frac{T}{2}}$ reduces to $-(\mathcal{M}_3^M+B_0 P_m)$.
The term $-\mathcal{M}_3^M$ thus cancels the corresponding term in $H_p$ so that undesired processes of the mechanical degree of freedom disappear in leading orders.
For the interaction terms, this cancellation is not perfect,
but there is a residual interaction $\omega_mk^3B_0 P_m/3$ in the effective Hamiltonian.
This term, however, is reduced by a factor of $3$, as compared to what can be achieved without intervals of free evolution.

The introduction of half-periods of undriven dynamics thus permits to improve the accuracy of desired effective Hamiltonians substantially.
It does, however, require a careful reconsideration of the discussion in Sec.~\ref{sec:EffHam},
in particular \eq{product} that is based on driving protocols in which the driving amplitude is changed after full periods of driving only.
Because $U_{\frac{T}{2}}$ commutes with the operator $V_{\phi_j}$ in \eq{productOfTwoPeriods}, the dynamics of any time-window of
two periods of driven dynamics
is still described by the propagator $U_d=\exp(-iH_pT)$ with $H_p$ defined in \eq{thirdOrderHamWithCorrPhase},
so that the discussion resulting in \eq{product} naturally includes the case of half-periods of undriven dynamics.

\begin{figure}[t!]
  \centering
  \includegraphics[keepaspectratio, width=0.5\textwidth]{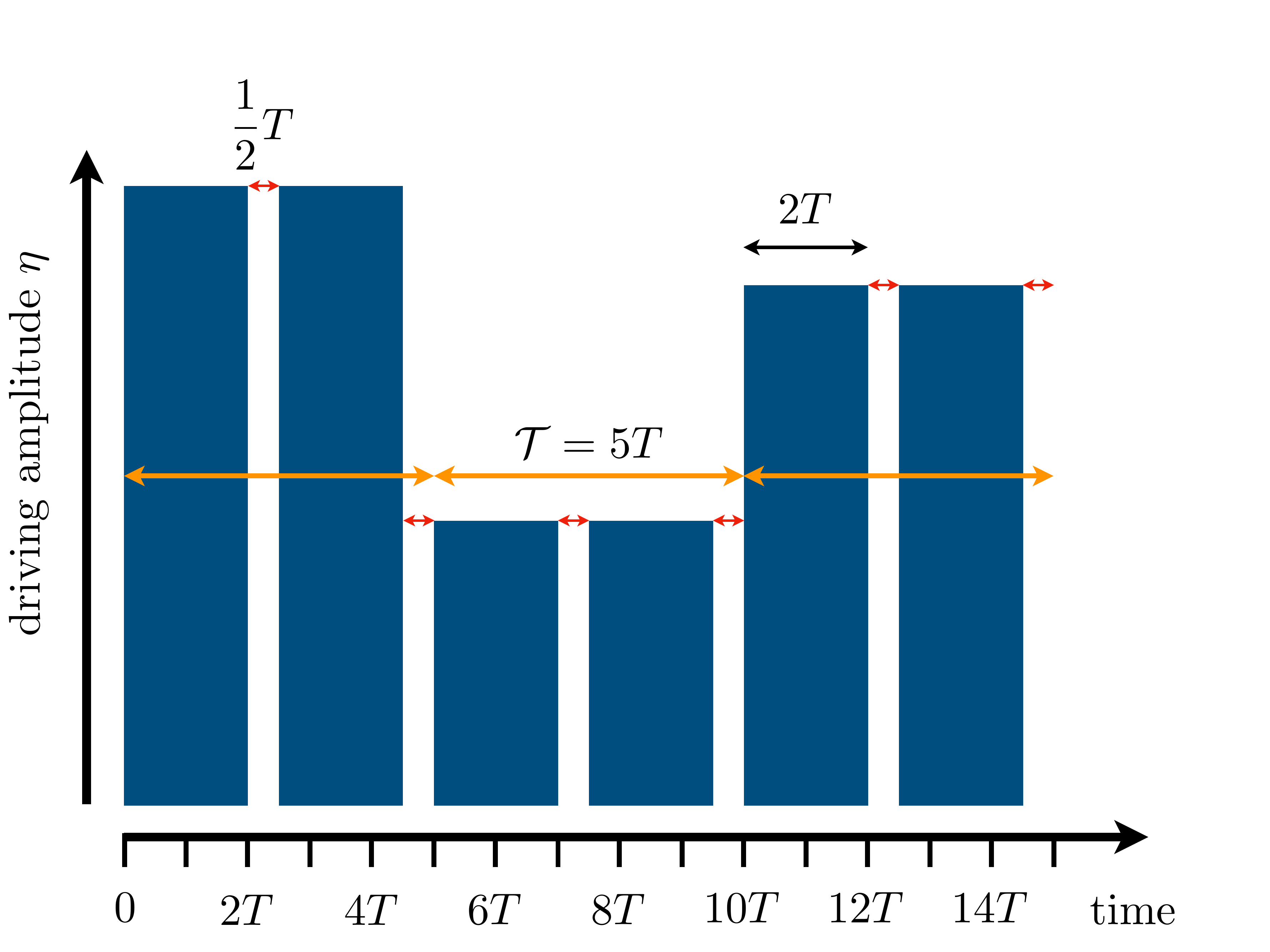}
  \caption{
  Schematic example of a driving pattern of duration $15T=3{\cal T}$.
  The protocol is comprised of three intervals of length ${\cal T}=5T$ (orange, long arrows) characterized by a constant value of the driving amplitude $\eta$.
  Each of these intervals consists of a pair of time-windows of length $2T$ of driven dynamics, each of which is followed by half a period of undriven dynamics (red, short arrows).
  }
  \label{fig:pattern} 
\end{figure}

With a pair of time-windows of length $2T$ of driven dynamics each of which is followed by half a period of undriven dynamics as sketched in Fig.~\ref{fig:pattern},
the effective Hamiltonian for the dynamics over $\fiveT=5T$ reads
\be\label{eq:finalEffectiveHam}
H=\omega_m\brkts{k^2H^{(2)}+k^3H^{(3)}}\new{\ ,}
\ee
with
\be\label{eq:finalEffectiveHamTerms}
H^{(2)}&=\frac{2}{3}\eta^2\brkts{a^2+(a^\dag)^2}-5n_c^2\new{\ ,} \\
H^{(3)}&=\frac{2}{3}\eta^2\brkts{(a^\dag)^2-a^2}\brkts{b-b^\dag}\new{\ .}
\ee
While the effective Hamiltonian given in \eq{effHam2ndFin} could have been taken as starting point for the subsequent analysis,
the suppression of undesired processes that is achieved with only a moderately more involved protocol, makes \eq{finalEffectiveHam} the preferred choice.

\subsection{Optimization}

With the driving profiles devised so far, it is ensured that undesired processes are largely suppressed,
and the amplitude of the driving field can be chosen in accordance with the state that is to be prepared.
In order to realise high fidelity state preparations, it is helpful to consider a series of several intervals $\ft$ of driven dynamics,
and to optimize over the driving amplitudes of each of those intervals.

Since such optimizations require the analysis of system dynamics with several different patterns of driving amplitudes, efficiency in the numerical propagation is essential.
We therefore use the effective Hamiltonian $H^{(2)}$ in \eq{finalEffectiveHamTerms}.
Since to this level of approximation, there is no interaction between the cavity field and the mechanical oscillator, this permits to restrict the numerical propagation to the dynamics of the cavity field only.

All the pulses discussed in Sec.\ref{sec:examples} are optimized  based on simulations with a Hilbert space for the cavity field that is truncated to the lowest $60$ Fock states,
and constraints on maximally admitted driving amplitude $\eta$.
These constrains ensure that the obtained solutions are compatible with practical constraints, and they help to avoid truncation errors.

\section{Optimized state preparation}
\label{sec:examples}

The framework developed in Sec.\ref{sec:theory} permits to identify driving patterns that result in the desired evolution towards nonclassical pure states of light in the regime of strong optomechanical coupling.
In subsection \ref{sec:coherentDym} we will discuss a range of achievable states and assess the validity of the perturbative approximation.
In subsection~\ref{sec:dissipative}, we will discuss the impact of dissipative effects on the achievable states.
All simulation results are computed using the Python toolbox Qutip~\cite{Qutip1,Qutip2}.

The accuracy of the state preparation will be assessed in terms of the fidelities
\be
F(\varrho,\rho)=
\mbox{Tr}
\sqrt{\sqrt{\rho}\varrho\sqrt{\rho}}\ ,
\ee
and
\be
F(\varrho,\ket{\Psi})=\sqrt{\bra{\Psi}\varrho\ket{\Psi}}
\ee
that specify the similarity between the state $\varrho$ and a mixed or pure state $\rho$ and $\ket{\Psi}$.

In order to discriminate between the limited accuracy of the perturbative expansion and the quality of the optimized driving profiles,
it will be helpful to define three different fidelities in terms of the numerically exact propagator $\Lambda_n$ of the coherent system dynamics, the propagator $\Lambda_l$ of the dissipative system dynamics, and the second order perturbative propagator $\Lambda_p$ of the system dynamics.

With the cavity and mechanical oscillator initialized in their ground state and when the system is lossless, the numerically exact final state of the cavity is given by
\be
\varrho_n=\mbox{Tr}_M\Lambda_n(\ket{0}\bra{0}\otimes\ket{0}\bra{0})\ ,
\ee
where the symbol $\mbox{Tr}_M$ denotes the trace over the mechanical degree of freedom.
Similarly, the final state in perturbative approximation reads
\be
\varrho_p=\mbox{Tr}_M\Lambda_p(\ket{0}\bra{0}\otimes\ket{0}\bra{0})\ .
\ee
Finally, when any system imperfection is involved, the numerically exact final state reads
\be
\varrho_l=\mbox{Tr}_M\Lambda_l(\ket{0}\bra{0}\otimes\ket{0}\bra{0})\ .
\ee
 
For any given target state $\ket{\Psi_t}$ of the cavity, we can thus define the fidelity
\be
F_n=F(\varrho_n,\ket{\Psi_t})
\ee
that specifies how well the goal of optimization is achieved in a lossless system, and the fidelity
\be
F_l=F(\varrho_l,\ket{\Psi_t})
\ee
that specifies how well the goal of optimization is achieved when relevant experimental noises are considered.
Lastly, the fidelity 
\be
F_i=F(\varrho_n,\varrho_l)
\ee
characterizes the impact of the relevant experimental noises with other imperfections isolated.

\subsection{Coherent dynamics}\label{sec:coherentDym}

\subsubsection{Fock state}\label{sec:fock}

Given the suitability of the present control scheme for the creation of photon pairs,
the creation of the Fock state $\ket{2}$ of the light field, starting from the cavity field and the oscillator in their ground state is a natural task.

\begin{figure}[t!]
  \centering
  \includegraphics[keepaspectratio, width=0.5\textwidth]{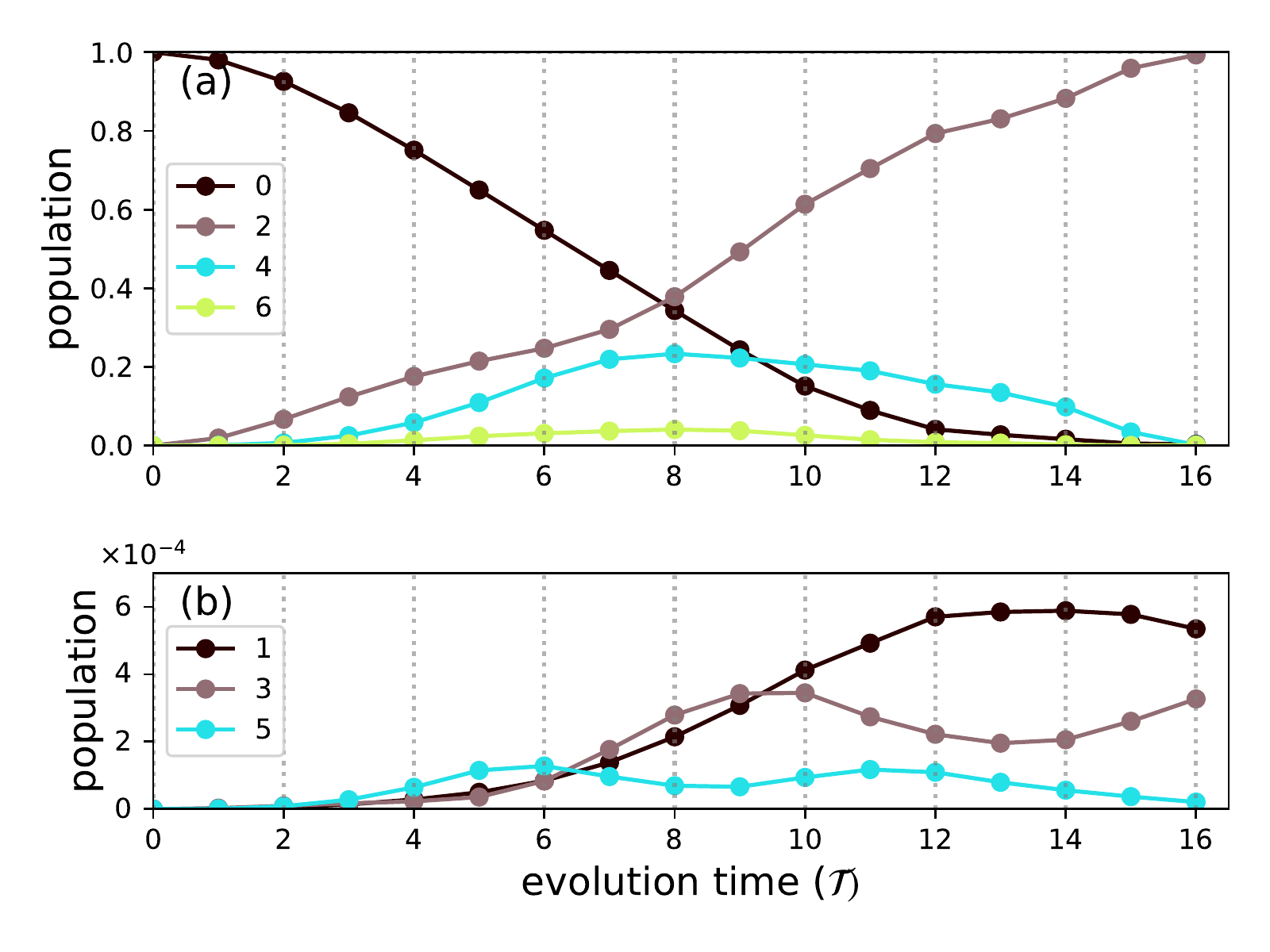}
  \caption{Plots of populations of Fock states in the cavity. Inset (a) plots populations of even Fock states and inset (b) plots populations of odd Fock states. The relative coupling strength is $k=1/26$, the relative driving strengths $\eta$ are optimized with a maximally admissible value $\eta_{max}=4$, and the length of evolution is $t_f=16\fiveT$. A data point is plotted after each five mechanical periods $\fiveT$. The final state is almost a Fock state $\ket{2}$.}   
  \label{fig:k26t16FockSimul} 
\end{figure}

\new{Fig.~\ref{fig:k26t16FockSimul}} depicts the dynamics of the cavity field under optimised driving for an evolution time of $16\fiveT$ with a maximum admissible driving strength $\eta_{max}=4$ and the coupling strength $k=1/26$.

Insets \new{(a) and (b)} depict the time-dependent occupation of the lowest $6$ Fock states.
Due to the suppression of the creation of single photons, discussed in Sec.\ref{sec:EffHam} and Sec.\ref{sec:drivingpattern},
the populations of odd Fock states remain orders of magnitudes smaller than the populations of even Fock states.
There is a sizeable population of the state $\ket{4}$ and $\ket{6}$ during the dynamics;
that is, despite the suppression of excitations to higher-lying states,
these states do become occupied.
The numerically optimized driving pattern, however, induces a dynamics in which these undesired states become un-occupied in the final state,
and a final fidelity $F_n=0.997$ is obtained.

In the idealized situation of lossless dynamics, one would expect to obtain best results in the limit of long evolution times with weak interactions ({\it i.e.} $k\ll 1)$, since this is the limit in which the underlying perturbative treatment becomes exact.
In practice, however, it is necessary to restrict the dynamics to a short time-window so that dissipative effects do not affect the state preparation too strongly.
Shorter evolution times will generally require stronger interactions, and too strong interactions can become conflicting with the perturbative approximation.

\begin{figure}[t!]
  \centering
  \includegraphics[keepaspectratio, width=0.5\textwidth]{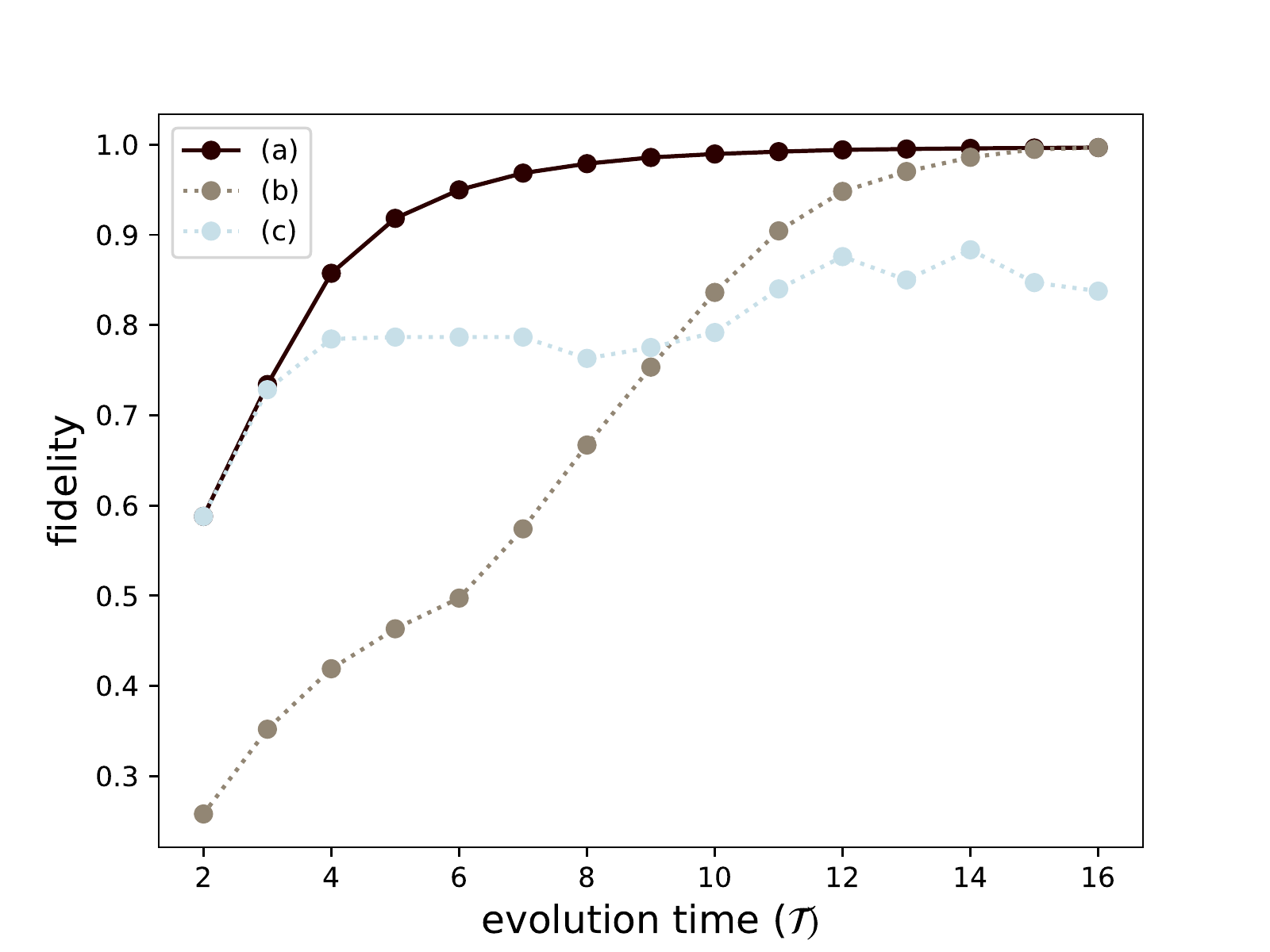}
  \caption{Plot of the fidelity $F_n$ against the evolution time in $\fiveT$ when $\eta_{max}=4$. (a) For each time point, the coupling strength $k$ is optimized to achieve the highest possible fidelity. (b) The coupling strength is fixed to be $k=1/26$. (c) The coupling strength is fixed to be $k=1/13$.
  Whether or not a suitable combination of interaction strength and evolution time is selected has a great impact on the final fidelity $F_n$.}   
  \label{fig:fidelVSk} 
\end{figure}

\begin{figure}[t!]
  \centering
  \includegraphics[keepaspectratio, width=0.5\textwidth]{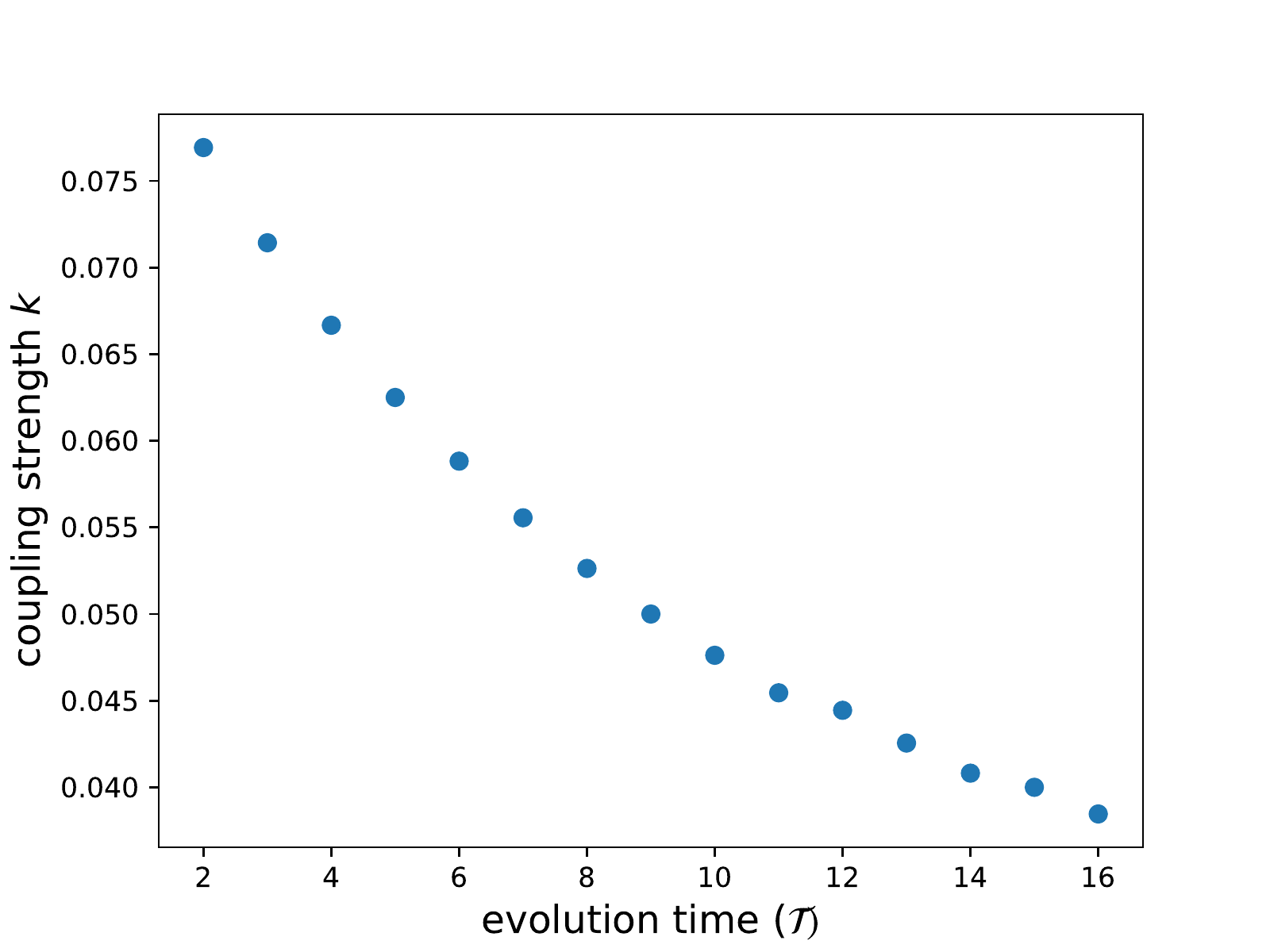}
  \caption{Plot of the coupling strength that is optimal for a given evolution time ranging from $2\fiveT$ to $16\fiveT$ when $\eta_{max}=4$. 
  Optimal coupling strength doubles when the evolution time is restricted from $16\fiveT$ to $2\fiveT$.
  }   
  \label{fig:kVSt} 
\end{figure}

For any given evolution time, one would thus expect to find an optimal value of the interaction constant $k$.
Fig.~\ref{fig:fidelVSk} depicts the fidelity $F_n$ (solid) obtained with optimized driving and the interaction strength that was found to be optimal for any given evolution time.
Dashed lines indicate the fidelity $F_n$ obtained with optimized driving profiles with fixed interaction strengths
$k=1/13$ and $k=1/26$
that are optimal for the evolution times $2\fiveT$ to $16\fiveT$.
Comparison of \new{dotted} and solid lines highlights the substantial gain in fidelity of state preparation that can be obtained by selecting a suitable combination of interaction strength and evolution time.
Fig.~\ref{fig:kVSt}
shows the coupling strength that is optimal for a given evolution time ranging from $2\fiveT$ to $16\fiveT$.
Even though the range of optimal coupling strength varies only by a factor of $2$,
the optimal choice of the interaction constant has a strong impact on the achievable state fidelities.
For example, a fidelity of $0.990$ can be achieved with an evolution time as short as $10\fiveT$ with the optimal interaction strengths, while an interaction strength of $k=1/26$ would only result in a fidelity of $0.836$ within the same time.

\subsubsection{superposition states}\label{sec:superpos}

Similarly to the creation of Fock states, the present framework can also be used to find driving patterns for the creation of coherent superpositions of Fock states.
This will be exemplified in the following with the target state
\be
\ket{\Psi_\vartheta}=\frac{1}{\sqrt{2}}\left(\ket{0}+e^{i\vartheta}\ket{2}\right)\ .
\ee
In addition to the optimization of the driving profile, the following optimization includes an optimization over the relative phase $\vartheta$,
{\it i.e.} it identifies the target state that is best suited among all balanced superpositions of the Fock state $\ket{0}$ and $\ket{2}$.

\begin{figure}[t!]
  \centering
  \includegraphics[keepaspectratio, width=0.5\textwidth]{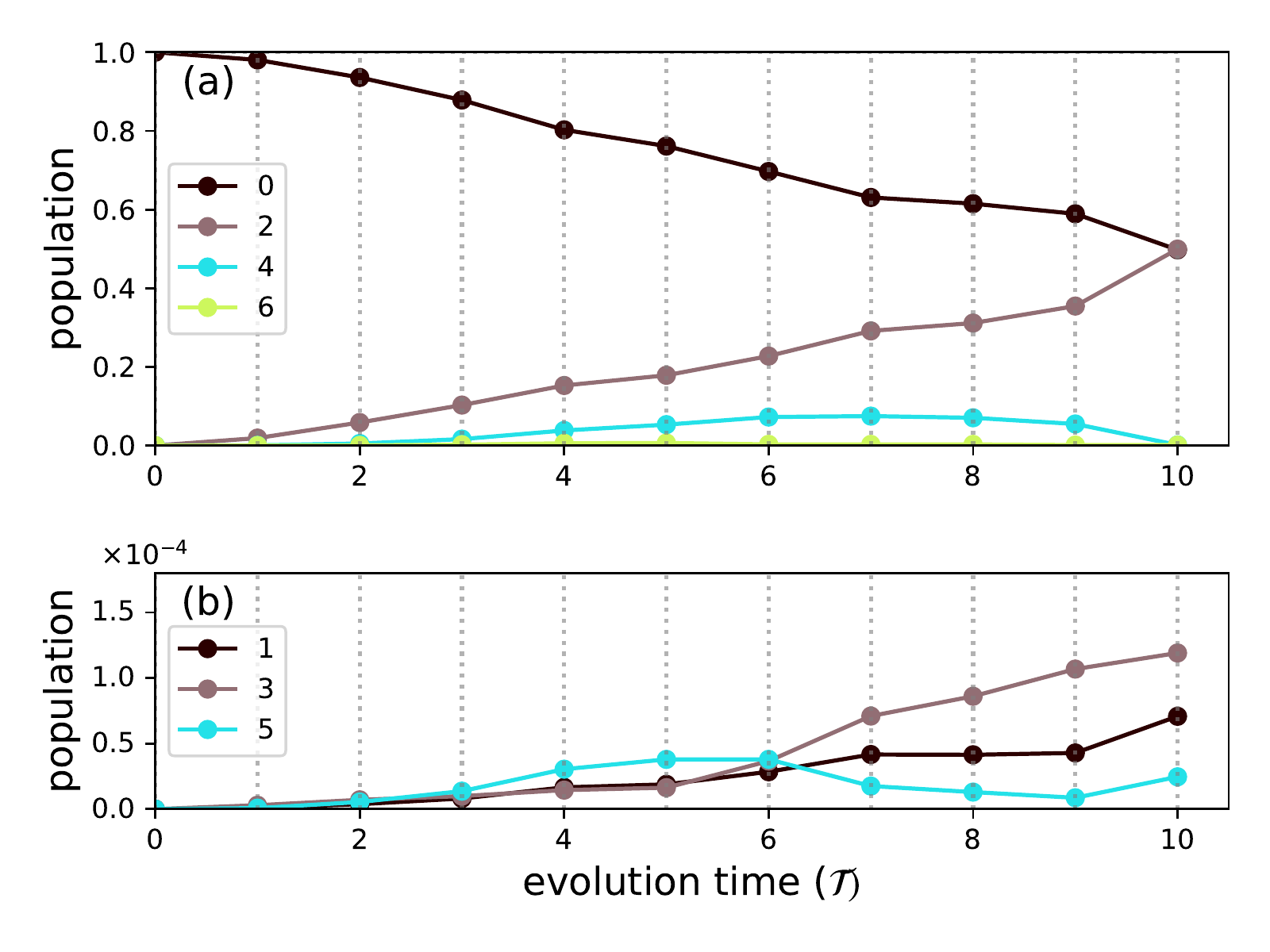}
  \caption{Plots of populations of Fock states in the cavity. Inset (a) plots populations of even Fock states and inset (b) plots populations of odd Fock states. The relative coupling strength $k=1/26$, the relative driving strengths $\eta$ are optimized with a maximally admissible value $\eta_{max}=4$, and the length of evolution is $t_f=10\fiveT$. A data point is plotted after each five mechanical periods $\fiveT$. The final state is almost an equal superposition state $1/\sqrt{2}\left(\ket{0}+e^{-1.86i}\ket{2}\right)$.}   
  \label{fig:k26t16SuperPosSimul} 
\end{figure}

Similar as in the case of Fock states, the driving strengths are optimized for a maximally admissible driving strength $\eta_{max}=4$ and coupling strength $k=1/26$.
However, because the expected number of photons in the cavity is less for an equal superposition state $\ket{\Psi_\vartheta}$ than for a Fock state $\ket{2}$, the evolution time can be reduced from $16\fiveT$ to $10\fiveT$ while keeping the fidelity as high as $F_n=0.998$.
The dynamics with given parameters is plotted in \new{Fig.~\ref{fig:k26t16SuperPosSimul}} which also shows negligible occupation of the Fock states that do not contribute to the desired superposition state.

\subsection{Dissipative dynamics}
\label{sec:dissipative}
The final question to be discussed is the impact of experimental noise
on the final state~\footnote{Simulations of dissipative dynamics are performed bases on the Trotter-Suzuki decomposition with terms induced by time-independent effective Hamiltonian and terms induced by Lindbladians for the dissipative part of the dynamics.}.
The two most significant experimental imperfections
are leakage of light from the cavity and thermalization in the mechanical oscillator.
The latter can result in thermal excitations in the initial state of the mechanical oscillator, and it can result in dissipative dynamics during the process of state preparation.

In order to analyse the impact of experimental noise, this section addresses the accuracy of state preparation under various imperfections.
This will be exemplified with the control pulses identified as optimal for the noiseless system and with the three optimal pairs $(1/26,16\fiveT)$, $(1/21,10\fiveT)$ and $(1/16,5\fiveT)$ of parameter values of coupling strength and evolution time. For all three parameter sets, the maximally admissible driving strength is fixed to be $\eta_{max}=4$.

\subsubsection{Thermal Initial States}
\begin{figure}[t!]
  \centering
  \includegraphics[keepaspectratio, width=0.5\textwidth]{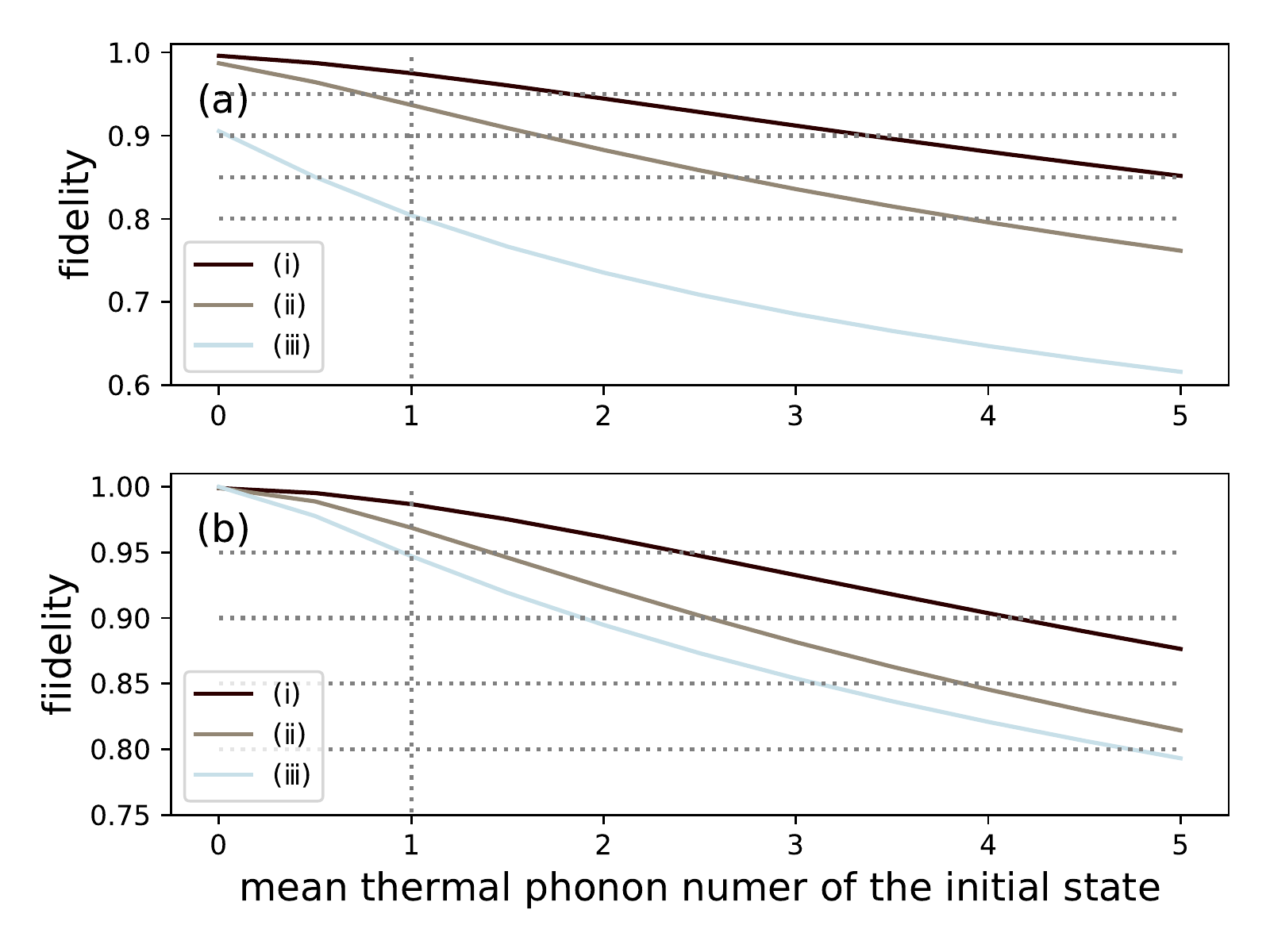}
  \caption{Plots of the fidelities (a) $F_l$ and (b) $F_i$ against mean thermal phonon numbers $\bar{n}_{th}$ of initial states. Curves \mbox{\RN{1},} \RN{2} and \RN{3} represent cases when parameters $(k,\eta_{max},t_f)$ equal to $(1/26,4,16\fiveT)$, $(1/21,4,10\fiveT)$ and $(1/16,4,5\fiveT)$ respectively.
All fidelities remain high when $\bar{n}_{th}<1$ except for the case of $(1/16,4,5\fiveT)$.
The exception is caused by the higher order entanglement that is amplified by excitations in the mechanical oscillator.}
  \label{fig:thermalInit} 
\end{figure}

In this subsection, we consider noiseless, unitary dynamics, but thermal initial state of the mechanical oscillator.
Fig.~\ref{fig:thermalInit} depicts the fidelities $F_l$ and $F_i$
as functions of the mean thermal phonon number of the initial mechanical state.
Fidelity $F_l$ assesses the overall accuracy of the state preparation, whereas $F_i$ isolates the impact the thermal excitations.
The difference between these two fidelities becomes best apparent in the case $(1/16,4,5\fiveT)$, in which $F_n$ does not reach the ideal value of unity for vanishing thermal excitations.

In addition to the fact that imperfections reduce the state fidelities most strongly for strong interactions and fast protocols,
also the impact of thermal excitations is most pronounced in these cases.

This can be explained by the fact that shorter dynamics require either stronger coupling strengths or stronger driving strengths, and the increase of either of the two parameters will lead to larger coefficients before undesired terms such as $k^3H^{(3)}$ in the perturbative solution in \eq{finalEffectiveHam}.
These undesired terms including entanglement between the cavity and the oscillator are further amplified by the non-vanishing phononic occupation, and thus will lead to lower final fidelities.

\subsubsection{Optical Loss}

Leakage of photons from the cavity at a rate $\kappa$ can be modelled with the Lindbladian
\be\label{eq:opLindOp}
{\cal L}\varrho=\kappa\mathcal{D}[a]\rho\equiv\kappa\left(a\varrho a^\dagger-\frac{1}{2}\{a^\dagger a,\varrho\}\right)\ ,
\ee
which, together with the system Hamiltonian defines a Master equation.

\begin{figure}[t!]
  \centering
  \includegraphics[keepaspectratio, width=0.5\textwidth]{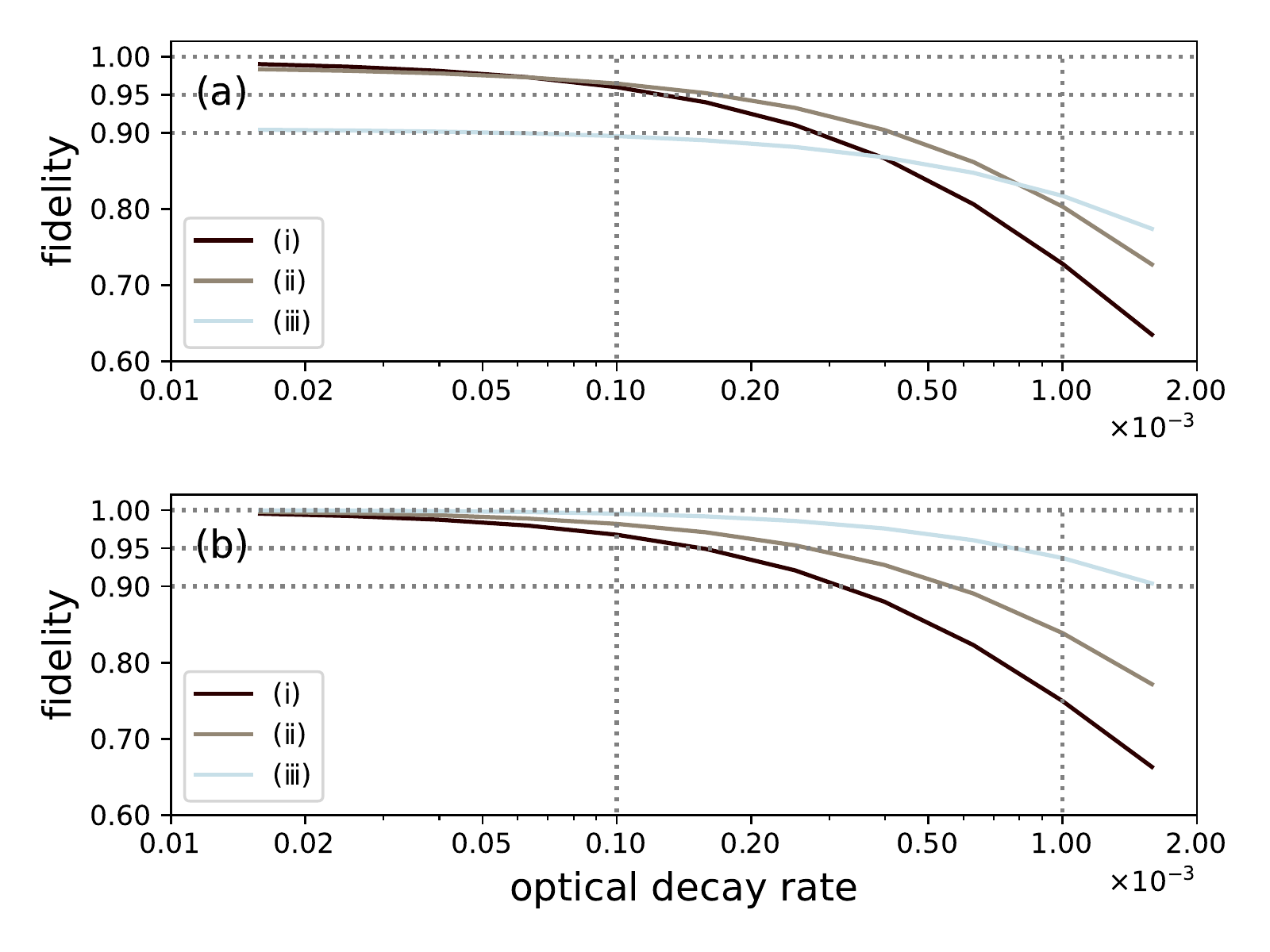}
  \caption{Plots of the fidelities (a) $F_l$ and (b) $F_i$ against relative optical decay rates $\kappa/\omega_m$ of the lossy cavity.
 Curves \mbox{\RN{1},} \RN{2} and \RN{3} represent cases when parameters $(k,\eta_{max},t_f)$ equal to $(1/26,4,16\fiveT)$, $(1/21,4,10\fiveT)$ and $(1/16,4,5\fiveT)$ respectively. 
The x-axis is in log scale while the y-axis is in linear scale. 
The plots suggest that the fidelity drops dramatically with the increase in optical decay rate when the decay rate exceeds $10^{-4}\omega_m$, and that shorter evolution time leads to a more resilient state.}
  \label{fig:opticalLoss} 
\end{figure}

Fig.~\ref{fig:opticalLoss} depicts the fidelities $F_l$ and $F_i$ as function of the optical decay rate $\kappa$.
Just like in Fig.~\ref{fig:thermalInit} the fidelities $F_l$ remain smaller than $1$ for $\kappa\to 0$;
but, in contrast to the case of initial thermal excitations, the faster protocols in systems with stronger interactions become favourable with stronger optical decay.
\new{Fig.~\ref{fig:opticalLoss} (a)} thus indicates at what values of $\kappa$ the coherent imperfections outweigh the incoherent imperfections,
and helps to identify the coupling strength and corresponding duration that is best adopted for a given level of optical loss.

In order to reach fidelities over $0.9$, the optical decay rate may at most be on the order of $10^{-4}\omega_m$.
The best value achieved by the current state-of-the-art~\cite{DGH19} in the strong coupling regime is still approximately two orders of magnitude away from the required value, but given the steady pace over the past decades, the regime may eventually be reached in the future.

\subsubsection{Mechanical Thermalization}

Finally, systems with vanishing optical decay rates but with finite mechanical decay rates $\gamma$ are considered.
The thermalization of the mechanical oscillator can be modelled with the Lindbladian~\cite{HHL+15}
\be\label{eq:meLindOp}
{\cal L}\varrho=&\gamma(\bar{n}_{b}+1)\mathcal{D}[b-ka^\dag a]\rho\\
&+\gamma\bar{n}_{b}\mathcal{D}[b^\dag-ka^\dag a]\rho\\
&+\frac{4\gamma k^2}{log(1+\frac{1}{\bar{n}_{b}})}\mathcal{D}[a^\dag a]\rho \ ,
\ee
with $\bar{n}_{b}$ being the mean thermal phonon number of the environment.
The unusual shift of the mechanical annihilation and creation operators depending on the photon number operator $a^\dag a$ and the dephasing term depending on the temperature of the oscillator result from the non-vanishing value of the coupling strength $k$.

\begin{figure}[t!]
  \centering
  \includegraphics[keepaspectratio, width=0.5\textwidth]{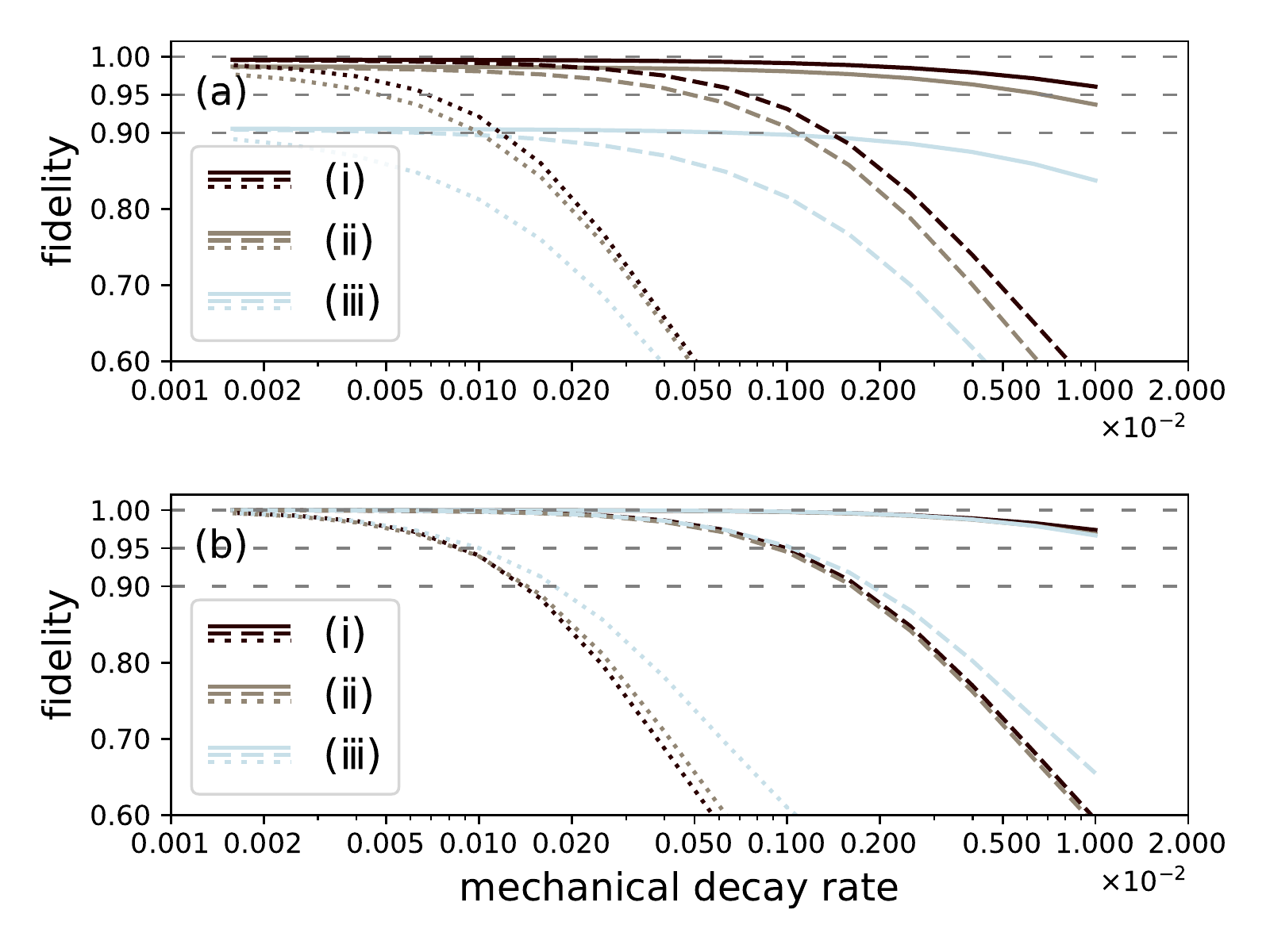}
  \caption{Plots of the fidelities (a) $F_l$ and (b) $F_i$ against relative mechanical decay rate $\gamma/\omega_m$ of the oscillator.
 Curves \mbox{\RN{1},} \RN{2} and \RN{3} represent cases when parameters $(k,\eta_{max},t_f)$ equal to $(1/26,4,16\fiveT)$, $(1/21,4,10\fiveT)$ and $(1/16,4,5\fiveT)$ respectively. 
 Solid, dashed, and dotted curves represent cases when the thermal bath contains on average $\bar{n}_{b}=1$, $\bar{n}_{b}=10$ and $\bar{n}_{b}=100$ phonons respectively.
The x-axis is in log scale and the y-axis is in linear scale.  
The plots suggest that the infidelity resulting from mechanical imperfection is negligible compared to that from optical imperfection when the thermal phonon number $\bar{n}_{b}\ll100$, but becomes noticeable otherwise.}
  \label{fig:mechanicalLoss} 
\end{figure}

The fidelities \new{$F_l$ and $F_i$} are plotted in Fig.~\ref{fig:mechanicalLoss} as functions of the mechanical decay rate $\gamma$.
The plot indicates that faster protocols which require stronger optomechanical interactions are slightly favourable for higher $F_i$ but the advantage is not significant when taking into the account the fact that faster protocols have lower $F_n$ \new{in the dissipation-less case.}

When the thermal bath contains $10$ or less phonons, effect of mechanical thermalization is completely negligible as compared to effect of optical decay when,
as observed experimentally~\cite{AKM14}, the mechanical decay rate $\gamma$ is smaller than the optical decay rate $\kappa$.
Especially when $\bar{n}_b\leq1$, a fidelity $F_i>0.95$ can be achieved with $\gamma<10^{-2}\omega_m$ which can be realized under several existing experimental setups~\cite{AKM14,DGH19}.
However, for systems that cannot maintain a low thermal phonon number $\bar{n}_b\ll100$, the mechanical decay $\gamma$ is required to be at least smaller than $10^{-3}\omega_m$ to reach a fidelity of $0.9$.

\section{Conclusions and Outlook}

The tools for state-preparation developed here, help to overcome the low success rates of probabilistic protocols.
Even though the non-linear interaction between light-field and mechanical oscillator tends to result in growing entanglement between the two degrees of freedom, the present driving patterns achieve close-to-perfect unitary dynamics of the light field while ensuring that the benefits of the non-linear interactions are preserved for the realization of non-Gaussian states.

The extra freedom of pulse shaping gives access to a variety of quantum states of light, beyond the two-photon Fock state and superposition states discussed here in more detail.
The high fidelities that can be obtained in the presence of optical loss and mechanical heating highlight the experimental feasibility of deterministic state preparation in upcoming generations of optomechanical experiments with increasing coupling strength between optical and mechanical components.

\new{While the driving protocols derived here are based on the coherent optomechanical dynamics only,
the formalism can be generalized to dissipative dynamics.
This would enable the optimization of the interplay of driving and environment engineering~\cite{KMC13,ACV+13,AV14} and to include the process of coupling light outside the cavity~\cite{CDG+10}.}

\section{Acknowledgement}

We are grateful for stimulating discussions with Daniel Burgarth, Myungshik Kim, Sean Greenaway and Jack Clarke.
Numerical simulations were carried out on Imperial HPC facilities~\cite{imperialHPC}.

\nocite{apsrev41Control}
\bibliographystyle{apsrev4-2}
\bibliography{opto}

\begin{thebibliography}{44}%
\makeatletter
\providecommand \@ifxundefined [1]{%
 \@ifx{#1\undefined}
}%
\providecommand \@ifnum [1]{%
 \ifnum #1\expandafter \@firstoftwo
 \else \expandafter \@secondoftwo
 \fi
}%
\providecommand \@ifx [1]{%
 \ifx #1\expandafter \@firstoftwo
 \else \expandafter \@secondoftwo
 \fi
}%
\providecommand \natexlab [1]{#1}%
\providecommand \enquote  [1]{``#1''}%
\providecommand \bibnamefont  [1]{#1}%
\providecommand \bibfnamefont [1]{#1}%
\providecommand \citenamefont [1]{#1}%
\providecommand \href@noop [0]{\@secondoftwo}%
\providecommand \href [0]{\begingroup \@sanitize@url \@href}%
\providecommand \@href[1]{\@@startlink{#1}\@@href}%
\providecommand \@@href[1]{\endgroup#1\@@endlink}%
\providecommand \@sanitize@url [0]{\catcode `\\12\catcode `\$12\catcode
  `\&12\catcode `\#12\catcode `\^12\catcode `\_12\catcode `\%12\relax}%
\providecommand \@@startlink[1]{}%
\providecommand \@@endlink[0]{}%
\providecommand \url  [0]{\begingroup\@sanitize@url \@url }%
\providecommand \@url [1]{\endgroup\@href {#1}{\urlprefix }}%
\providecommand \urlprefix  [0]{URL }%
\providecommand \Eprint [0]{\href }%
\providecommand \doibase [0]{https://doi.org/}%
\providecommand \selectlanguage [0]{\@gobble}%
\providecommand \bibinfo  [0]{\@secondoftwo}%
\providecommand \bibfield  [0]{\@secondoftwo}%
\providecommand \translation [1]{[#1]}%
\providecommand \BibitemOpen [0]{}%
\providecommand \bibitemStop [0]{}%
\providecommand \bibitemNoStop [0]{.\EOS\space}%
\providecommand \EOS [0]{\spacefactor3000\relax}%
\providecommand \BibitemShut  [1]{\csname bibitem#1\endcsname}%
\let\auto@bib@innerbib\@empty
\bibitem [{\citenamefont {Aspelmeyer}\ \emph {et~al.}(2014)\citenamefont
  {Aspelmeyer}, \citenamefont {Kippenberg},\ and\ \citenamefont
  {Marquardt}}]{AKM14}%
  \BibitemOpen
  \bibfield  {author} {\bibinfo {author} {\bibfnamefont {M.}~\bibnamefont
  {Aspelmeyer}}, \bibinfo {author} {\bibfnamefont {T.~J.}\ \bibnamefont
  {Kippenberg}},\ and\ \bibinfo {author} {\bibfnamefont {F.}~\bibnamefont
  {Marquardt}},\ }\bibfield  {title} {\bibinfo {title} {Cavity optomechanics},\
  }\href {https://doi.org/10.1103/revmodphys.86.1391} {\bibfield  {journal}
  {\bibinfo  {journal} {Rev. Mod. Phys.}\ }\textbf {\bibinfo {volume} {86}},\
  \bibinfo {pages} {1391--1452} (\bibinfo {year} {2014})}\BibitemShut {NoStop}%
\bibitem [{\citenamefont {Khalili}\ \emph {et~al.}(2010)\citenamefont
  {Khalili}, \citenamefont {Danilishin}, \citenamefont {Miao}, \citenamefont
  {M\"{u}ller-Ebhardt}, \citenamefont {Yang},\ and\ \citenamefont
  {Chen}}]{KDM+10}%
  \BibitemOpen
  \bibfield  {author} {\bibinfo {author} {\bibfnamefont {F.}~\bibnamefont
  {Khalili}}, \bibinfo {author} {\bibfnamefont {S.}~\bibnamefont {Danilishin}},
  \bibinfo {author} {\bibfnamefont {H.}~\bibnamefont {Miao}}, \bibinfo {author}
  {\bibfnamefont {H.}~\bibnamefont {M\"{u}ller-Ebhardt}}, \bibinfo {author}
  {\bibfnamefont {H.}~\bibnamefont {Yang}},\ and\ \bibinfo {author}
  {\bibfnamefont {Y.}~\bibnamefont {Chen}},\ }\bibfield  {title} {\bibinfo
  {title} {Preparing a mechanical oscillator in non-gaussian quantum states},\
  }\href {https://doi.org/10.1103/physrevlett.105.070403} {\bibfield  {journal}
  {\bibinfo  {journal} {Phys. Rev. Lett.}\ }\textbf {\bibinfo {volume} {105}},\
  \bibinfo {pages} {070403} (\bibinfo {year} {2010})}\BibitemShut {NoStop}%
\bibitem [{\citenamefont {Vanner}\ \emph {et~al.}(2011)\citenamefont {Vanner},
  \citenamefont {Pikovski}, \citenamefont {Cole}, \citenamefont {Kim},
  \citenamefont {Brukner}, \citenamefont {Hammerer}, \citenamefont {Milburn},\
  and\ \citenamefont {Aspelmeyer}}]{VPC+11}%
  \BibitemOpen
  \bibfield  {author} {\bibinfo {author} {\bibfnamefont {M.~R.}\ \bibnamefont
  {Vanner}}, \bibinfo {author} {\bibfnamefont {I.}~\bibnamefont {Pikovski}},
  \bibinfo {author} {\bibfnamefont {G.~D.}\ \bibnamefont {Cole}}, \bibinfo
  {author} {\bibfnamefont {M.~S.}\ \bibnamefont {Kim}}, \bibinfo {author}
  {\bibfnamefont {C.}~\bibnamefont {Brukner}}, \bibinfo {author} {\bibfnamefont
  {K.}~\bibnamefont {Hammerer}}, \bibinfo {author} {\bibfnamefont {G.~J.}\
  \bibnamefont {Milburn}},\ and\ \bibinfo {author} {\bibfnamefont
  {M.}~\bibnamefont {Aspelmeyer}},\ }\bibfield  {title} {\bibinfo {title}
  {Pulsed quantum optomechanics},\ }\href
  {https://doi.org/10.1073/pnas.1105098108} {\bibfield  {journal} {\bibinfo
  {journal} {Proc. Natl. Acad. Sci. U.S.A.}\ }\textbf {\bibinfo {volume}
  {108}},\ \bibinfo {pages} {16182--16187} (\bibinfo {year}
  {2011})}\BibitemShut {NoStop}%
\bibitem [{\citenamefont {Rabl}(2011)}]{Rab11}%
  \BibitemOpen
  \bibfield  {author} {\bibinfo {author} {\bibfnamefont {P.}~\bibnamefont
  {Rabl}},\ }\bibfield  {title} {\bibinfo {title} {Photon blockade effect in
  optomechanical systems},\ }\href
  {https://doi.org/10.1103/physrevlett.107.063601} {\bibfield  {journal}
  {\bibinfo  {journal} {Phys. Rev. Lett.}\ }\textbf {\bibinfo {volume} {107}},\
  \bibinfo {pages} {063601} (\bibinfo {year} {2011})}\BibitemShut {NoStop}%
\bibitem [{\citenamefont {Brooks}\ \emph {et~al.}(2012)\citenamefont {Brooks},
  \citenamefont {Botter}, \citenamefont {Schreppler}, \citenamefont {Purdy},
  \citenamefont {Brahms},\ and\ \citenamefont {Stamper-Kurn}}]{BBS+12}%
  \BibitemOpen
  \bibfield  {author} {\bibinfo {author} {\bibfnamefont {D.~W.~C.}\
  \bibnamefont {Brooks}}, \bibinfo {author} {\bibfnamefont {T.}~\bibnamefont
  {Botter}}, \bibinfo {author} {\bibfnamefont {S.}~\bibnamefont {Schreppler}},
  \bibinfo {author} {\bibfnamefont {T.~P.}\ \bibnamefont {Purdy}}, \bibinfo
  {author} {\bibfnamefont {N.}~\bibnamefont {Brahms}},\ and\ \bibinfo {author}
  {\bibfnamefont {D.~M.}\ \bibnamefont {Stamper-Kurn}},\ }\bibfield  {title}
  {\bibinfo {title} {Non-classical light generated by quantum-noise-driven
  cavity optomechanics},\ }\href {https://doi.org/10.1038/nature11325}
  {\bibfield  {journal} {\bibinfo  {journal} {Nature}\ }\textbf {\bibinfo
  {volume} {488}},\ \bibinfo {pages} {476--480} (\bibinfo {year}
  {2012})}\BibitemShut {NoStop}%
\bibitem [{\citenamefont {Latmiral}\ and\ \citenamefont
  {Mintert}(2018)}]{LM18}%
  \BibitemOpen
  \bibfield  {author} {\bibinfo {author} {\bibfnamefont {L.}~\bibnamefont
  {Latmiral}}\ and\ \bibinfo {author} {\bibfnamefont {F.}~\bibnamefont
  {Mintert}},\ }\bibfield  {title} {\bibinfo {title} {Deterministic preparation
  of highly non-classical macroscopic quantum states},\ }\bibfield  {journal}
  {\bibinfo  {journal} {npj Quantum Inf.}\ }\textbf {\bibinfo {volume} {4}},\
  \href {https://doi.org/10.1038/s41534-018-0093-z} {10.1038/s41534-018-0093-z}
  (\bibinfo {year} {2018})\BibitemShut {NoStop}%
\bibitem [{\citenamefont {Xie}\ \emph {et~al.}(2019)\citenamefont {Xie},
  \citenamefont {Shang}, \citenamefont {Liao}, \citenamefont {Chen},\ and\
  \citenamefont {Lin}}]{XSL+19}%
  \BibitemOpen
  \bibfield  {author} {\bibinfo {author} {\bibfnamefont {H.}~\bibnamefont
  {Xie}}, \bibinfo {author} {\bibfnamefont {X.}~\bibnamefont {Shang}}, \bibinfo
  {author} {\bibfnamefont {C.-G.}\ \bibnamefont {Liao}}, \bibinfo {author}
  {\bibfnamefont {Z.-H.}\ \bibnamefont {Chen}},\ and\ \bibinfo {author}
  {\bibfnamefont {X.-M.}\ \bibnamefont {Lin}},\ }\bibfield  {title} {\bibinfo
  {title} {Macroscopic superposition states of a mechanical oscillator in an
  optomechanical system with quadratic coupling},\ }\href
  {https://doi.org/10.1103/physreva.100.033803} {\bibfield  {journal} {\bibinfo
   {journal} {Phys. Rev. A}\ }\textbf {\bibinfo {volume} {100}},\ \bibinfo
  {pages} {033803} (\bibinfo {year} {2019})}\BibitemShut {NoStop}%
\bibitem [{\citenamefont {Hu}\ \emph {et~al.}(2014)\citenamefont {Hu},
  \citenamefont {Wei}, \citenamefont {Huang},\ and\ \citenamefont
  {Liu}}]{HWH+14}%
  \BibitemOpen
  \bibfield  {author} {\bibinfo {author} {\bibfnamefont {L.-Y.}\ \bibnamefont
  {Hu}}, \bibinfo {author} {\bibfnamefont {C.-P.}\ \bibnamefont {Wei}},
  \bibinfo {author} {\bibfnamefont {J.-H.}\ \bibnamefont {Huang}},\ and\
  \bibinfo {author} {\bibfnamefont {C.-J.}\ \bibnamefont {Liu}},\ }\bibfield
  {title} {\bibinfo {title} {Quantum metrology with fock and even coherent
  states: Parity detection approaches to the heisenberg limit},\ }\href
  {https://doi.org/10.1016/j.optcom.2014.02.069} {\bibfield  {journal}
  {\bibinfo  {journal} {Opt. Commun.}\ }\textbf {\bibinfo {volume} {323}},\
  \bibinfo {pages} {68--76} (\bibinfo {year} {2014})}\BibitemShut {NoStop}%
\bibitem [{\citenamefont {Huver}\ \emph {et~al.}(2008)\citenamefont {Huver},
  \citenamefont {Wildfeuer},\ and\ \citenamefont {Dowling}}]{HWD08}%
  \BibitemOpen
  \bibfield  {author} {\bibinfo {author} {\bibfnamefont {S.~D.}\ \bibnamefont
  {Huver}}, \bibinfo {author} {\bibfnamefont {C.~F.}\ \bibnamefont
  {Wildfeuer}},\ and\ \bibinfo {author} {\bibfnamefont {J.~P.}\ \bibnamefont
  {Dowling}},\ }\bibfield  {title} {\bibinfo {title} {Entangled fock states for
  robust quantum optical metrology, imaging, and sensing},\ }\href
  {https://doi.org/10.1103/physreva.78.063828} {\bibfield  {journal} {\bibinfo
  {journal} {Phys. Rev. A}\ }\textbf {\bibinfo {volume} {78}},\ \bibinfo
  {pages} {063828} (\bibinfo {year} {2008})}\BibitemShut {NoStop}%
\bibitem [{\citenamefont {Demkowicz-Dobrza{\'{n}}ski}\ \emph
  {et~al.}(2015)\citenamefont {Demkowicz-Dobrza{\'{n}}ski}, \citenamefont
  {Jarzyna},\ and\ \citenamefont {Ko{\l}ody{\'{n}}ski}}]{DJK15}%
  \BibitemOpen
  \bibfield  {author} {\bibinfo {author} {\bibfnamefont {R.}~\bibnamefont
  {Demkowicz-Dobrza{\'{n}}ski}}, \bibinfo {author} {\bibfnamefont
  {M.}~\bibnamefont {Jarzyna}},\ and\ \bibinfo {author} {\bibfnamefont
  {J.}~\bibnamefont {Ko{\l}ody{\'{n}}ski}},\ }\bibfield  {title} {\bibinfo
  {title} {Quantum limits in optical interferometry},\ }in\ \href
  {https://doi.org/10.1016/bs.po.2015.02.003} {\emph {\bibinfo {booktitle}
  {Progress in Optics}}}\ (\bibinfo  {publisher} {Elsevier},\ \bibinfo {year}
  {2015})\ pp.\ \bibinfo {pages} {345--435}\BibitemShut {NoStop}%
\bibitem [{\citenamefont {Gisin}\ \emph {et~al.}(2002)\citenamefont {Gisin},
  \citenamefont {Ribordy}, \citenamefont {Tittel},\ and\ \citenamefont
  {Zbinden}}]{GRT+02}%
  \BibitemOpen
  \bibfield  {author} {\bibinfo {author} {\bibfnamefont {N.}~\bibnamefont
  {Gisin}}, \bibinfo {author} {\bibfnamefont {G.}~\bibnamefont {Ribordy}},
  \bibinfo {author} {\bibfnamefont {W.}~\bibnamefont {Tittel}},\ and\ \bibinfo
  {author} {\bibfnamefont {H.}~\bibnamefont {Zbinden}},\ }\bibfield  {title}
  {\bibinfo {title} {Quantum cryptography},\ }\href
  {https://doi.org/10.1103/revmodphys.74.145} {\bibfield  {journal} {\bibinfo
  {journal} {Rev. Mod. Phys.}\ }\textbf {\bibinfo {volume} {74}},\ \bibinfo
  {pages} {145--195} (\bibinfo {year} {2002})}\BibitemShut {NoStop}%
\bibitem [{\citenamefont {Yin}\ \emph {et~al.}(2010)\citenamefont {Yin},
  \citenamefont {Li}, \citenamefont {Chen}, \citenamefont {Han},\ and\
  \citenamefont {Guo}}]{YLC+10}%
  \BibitemOpen
  \bibfield  {author} {\bibinfo {author} {\bibfnamefont {Z.-Q.}\ \bibnamefont
  {Yin}}, \bibinfo {author} {\bibfnamefont {H.-W.}\ \bibnamefont {Li}},
  \bibinfo {author} {\bibfnamefont {W.}~\bibnamefont {Chen}}, \bibinfo {author}
  {\bibfnamefont {Z.-F.}\ \bibnamefont {Han}},\ and\ \bibinfo {author}
  {\bibfnamefont {G.-C.}\ \bibnamefont {Guo}},\ }\bibfield  {title} {\bibinfo
  {title} {Security of counterfactual quantum cryptography},\ }\href
  {https://doi.org/10.1103/physreva.82.042335} {\bibfield  {journal} {\bibinfo
  {journal} {Phys. Rev. A}\ }\textbf {\bibinfo {volume} {82}},\ \bibinfo
  {pages} {042335} (\bibinfo {year} {2010})}\BibitemShut {NoStop}%
\bibitem [{\citenamefont {Ac{\'{\i}}n}\ \emph {et~al.}(2009)\citenamefont
  {Ac{\'{\i}}n}, \citenamefont {Cerf}, \citenamefont {Ferraro},\ and\
  \citenamefont {Niset}}]{ACF+09}%
  \BibitemOpen
  \bibfield  {author} {\bibinfo {author} {\bibfnamefont {A.}~\bibnamefont
  {Ac{\'{\i}}n}}, \bibinfo {author} {\bibfnamefont {N.~J.}\ \bibnamefont
  {Cerf}}, \bibinfo {author} {\bibfnamefont {A.}~\bibnamefont {Ferraro}},\ and\
  \bibinfo {author} {\bibfnamefont {J.}~\bibnamefont {Niset}},\ }\bibfield
  {title} {\bibinfo {title} {Tests of multimode quantum nonlocality with
  homodyne measurements},\ }\href {https://doi.org/10.1103/physreva.79.012112}
  {\bibfield  {journal} {\bibinfo  {journal} {Phys. Rev. A}\ }\textbf {\bibinfo
  {volume} {79}},\ \bibinfo {pages} {012112} (\bibinfo {year}
  {2009})}\BibitemShut {NoStop}%
\bibitem [{\citenamefont {Menicucci}\ \emph {et~al.}(2006)\citenamefont
  {Menicucci}, \citenamefont {van Loock}, \citenamefont {Gu}, \citenamefont
  {Weedbrook}, \citenamefont {Ralph},\ and\ \citenamefont {Nielsen}}]{MLG+06}%
  \BibitemOpen
  \bibfield  {author} {\bibinfo {author} {\bibfnamefont {N.~C.}\ \bibnamefont
  {Menicucci}}, \bibinfo {author} {\bibfnamefont {P.}~\bibnamefont {van
  Loock}}, \bibinfo {author} {\bibfnamefont {M.}~\bibnamefont {Gu}}, \bibinfo
  {author} {\bibfnamefont {C.}~\bibnamefont {Weedbrook}}, \bibinfo {author}
  {\bibfnamefont {T.~C.}\ \bibnamefont {Ralph}},\ and\ \bibinfo {author}
  {\bibfnamefont {M.~A.}\ \bibnamefont {Nielsen}},\ }\bibfield  {title}
  {\bibinfo {title} {Universal quantum computation with continuous-variable
  cluster states},\ }\href {https://doi.org/10.1103/physrevlett.97.110501}
  {\bibfield  {journal} {\bibinfo  {journal} {Phys. Rev. Lett.}\ }\textbf
  {\bibinfo {volume} {97}},\ \bibinfo {pages} {110501} (\bibinfo {year}
  {2006})}\BibitemShut {NoStop}%
\bibitem [{\citenamefont {Niset}\ \emph {et~al.}(2009)\citenamefont {Niset},
  \citenamefont {Fiur{\'{a}}{\v{s}}ek},\ and\ \citenamefont {Cerf}}]{NFC09}%
  \BibitemOpen
  \bibfield  {author} {\bibinfo {author} {\bibfnamefont {J.}~\bibnamefont
  {Niset}}, \bibinfo {author} {\bibfnamefont {J.}~\bibnamefont
  {Fiur{\'{a}}{\v{s}}ek}},\ and\ \bibinfo {author} {\bibfnamefont {N.~J.}\
  \bibnamefont {Cerf}},\ }\bibfield  {title} {\bibinfo {title} {No-go theorem
  for gaussian quantum error correction},\ }\href
  {https://doi.org/10.1103/physrevlett.102.120501} {\bibfield  {journal}
  {\bibinfo  {journal} {Phys. Rev. Lett.}\ }\textbf {\bibinfo {volume} {102}},\
  \bibinfo {pages} {120501} (\bibinfo {year} {2009})}\BibitemShut {NoStop}%
\bibitem [{\citenamefont {Dell'Anno}\ \emph {et~al.}(2010)\citenamefont
  {Dell'Anno}, \citenamefont {De~Siena}, \citenamefont {Adesso},\ and\
  \citenamefont {Illuminati}}]{DSA+10}%
  \BibitemOpen
  \bibfield  {author} {\bibinfo {author} {\bibfnamefont {F.}~\bibnamefont
  {Dell'Anno}}, \bibinfo {author} {\bibfnamefont {S.}~\bibnamefont {De~Siena}},
  \bibinfo {author} {\bibfnamefont {G.}~\bibnamefont {Adesso}},\ and\ \bibinfo
  {author} {\bibfnamefont {F.}~\bibnamefont {Illuminati}},\ }\bibfield  {title}
  {\bibinfo {title} {Teleportation of squeezing: Optimization using
  non-gaussian resources},\ }\href {https://doi.org/10.1103/physreva.82.062329}
  {\bibfield  {journal} {\bibinfo  {journal} {Phys. Rev. A}\ }\textbf {\bibinfo
  {volume} {82}},\ \bibinfo {pages} {062329} (\bibinfo {year}
  {2010})}\BibitemShut {NoStop}%
\bibitem [{\citenamefont {Zhuang}\ \emph {et~al.}(2018)\citenamefont {Zhuang},
  \citenamefont {Shor},\ and\ \citenamefont {Shapiro}}]{ZSS18}%
  \BibitemOpen
  \bibfield  {author} {\bibinfo {author} {\bibfnamefont {Q.}~\bibnamefont
  {Zhuang}}, \bibinfo {author} {\bibfnamefont {P.~W.}\ \bibnamefont {Shor}},\
  and\ \bibinfo {author} {\bibfnamefont {J.~H.}\ \bibnamefont {Shapiro}},\
  }\bibfield  {title} {\bibinfo {title} {Resource theory of non-gaussian
  operations},\ }\href {https://doi.org/10.1103/physreva.97.052317} {\bibfield
  {journal} {\bibinfo  {journal} {Phys. Rev. A}\ }\textbf {\bibinfo {volume}
  {97}},\ \bibinfo {pages} {052317} (\bibinfo {year} {2018})}\BibitemShut
  {NoStop}%
\bibitem [{\citenamefont {Lemonde}\ \emph {et~al.}(2016)\citenamefont
  {Lemonde}, \citenamefont {Didier},\ and\ \citenamefont {Clerk}}]{LDC16}%
  \BibitemOpen
  \bibfield  {author} {\bibinfo {author} {\bibfnamefont {M.-A.}\ \bibnamefont
  {Lemonde}}, \bibinfo {author} {\bibfnamefont {N.}~\bibnamefont {Didier}},\
  and\ \bibinfo {author} {\bibfnamefont {A.~A.}\ \bibnamefont {Clerk}},\
  }\bibfield  {title} {\bibinfo {title} {Enhanced nonlinear interactions in
  quantum optomechanics via mechanical amplification},\ }\bibfield  {journal}
  {\bibinfo  {journal} {Nat. Commun.}\ }\textbf {\bibinfo {volume} {7}},\ \href
  {https://doi.org/10.1038/ncomms11338} {10.1038/ncomms11338} (\bibinfo {year}
  {2016})\BibitemShut {NoStop}%
\bibitem [{\citenamefont {Sarma}\ and\ \citenamefont {Sarma}(2018)}]{SS18}%
  \BibitemOpen
  \bibfield  {author} {\bibinfo {author} {\bibfnamefont {B.}~\bibnamefont
  {Sarma}}\ and\ \bibinfo {author} {\bibfnamefont {A.~K.}\ \bibnamefont
  {Sarma}},\ }\bibfield  {title} {\bibinfo {title} {Single-photon blockade in a
  hybrid cavity-optomechanical system via third-order nonlinearity},\ }\href
  {https://doi.org/10.1088/1361-6455/aab194} {\bibfield  {journal} {\bibinfo
  {journal} {J. Phys. B: At., Mol. Opt. Phys.}\ }\textbf {\bibinfo {volume}
  {51}},\ \bibinfo {pages} {075505} (\bibinfo {year} {2018})}\BibitemShut
  {NoStop}%
\bibitem [{\citenamefont {Shi}\ \emph {et~al.}(2018)\citenamefont {Shi},
  \citenamefont {Zhou}, \citenamefont {Xu},\ and\ \citenamefont
  {Liu}}]{SZX+18}%
  \BibitemOpen
  \bibfield  {author} {\bibinfo {author} {\bibfnamefont {H.-Q.}\ \bibnamefont
  {Shi}}, \bibinfo {author} {\bibfnamefont {X.-T.}\ \bibnamefont {Zhou}},
  \bibinfo {author} {\bibfnamefont {X.-W.}\ \bibnamefont {Xu}},\ and\ \bibinfo
  {author} {\bibfnamefont {N.-H.}\ \bibnamefont {Liu}},\ }\bibfield  {title}
  {\bibinfo {title} {Tunable phonon blockade in quadratically coupled
  optomechanical systems},\ }\bibfield  {journal} {\bibinfo  {journal} {Sci.
  Rep.}\ }\textbf {\bibinfo {volume} {8}},\ \href
  {https://doi.org/10.1038/s41598-018-20568-x} {10.1038/s41598-018-20568-x}
  (\bibinfo {year} {2018})\BibitemShut {NoStop}%
\bibitem [{\citenamefont {Wang}\ \emph {et~al.}(2019)\citenamefont {Wang},
  \citenamefont {Bai}, \citenamefont {Liu}, \citenamefont {Zhang},\ and\
  \citenamefont {Wang}}]{WBL+19}%
  \BibitemOpen
  \bibfield  {author} {\bibinfo {author} {\bibfnamefont {D.-Y.}\ \bibnamefont
  {Wang}}, \bibinfo {author} {\bibfnamefont {C.-H.}\ \bibnamefont {Bai}},
  \bibinfo {author} {\bibfnamefont {S.}~\bibnamefont {Liu}}, \bibinfo {author}
  {\bibfnamefont {S.}~\bibnamefont {Zhang}},\ and\ \bibinfo {author}
  {\bibfnamefont {H.-F.}\ \bibnamefont {Wang}},\ }\bibfield  {title} {\bibinfo
  {title} {Distinguishing photon blockade in a {PT} -symmetric optomechanical
  system},\ }\href {https://doi.org/10.1103/physreva.99.043818} {\bibfield
  {journal} {\bibinfo  {journal} {Phys. Rev. A}\ }\textbf {\bibinfo {volume}
  {99}},\ \bibinfo {pages} {043818} (\bibinfo {year} {2019})}\BibitemShut
  {NoStop}%
\bibitem [{\citenamefont {Wang}\ \emph {et~al.}(2020)\citenamefont {Wang},
  \citenamefont {Bai}, \citenamefont {Han}, \citenamefont {Liu}, \citenamefont
  {Zhang},\ and\ \citenamefont {Wang}}]{WBH+20}%
  \BibitemOpen
  \bibfield  {author} {\bibinfo {author} {\bibfnamefont {D.-Y.}\ \bibnamefont
  {Wang}}, \bibinfo {author} {\bibfnamefont {C.-H.}\ \bibnamefont {Bai}},
  \bibinfo {author} {\bibfnamefont {X.}~\bibnamefont {Han}}, \bibinfo {author}
  {\bibfnamefont {S.}~\bibnamefont {Liu}}, \bibinfo {author} {\bibfnamefont
  {S.}~\bibnamefont {Zhang}},\ and\ \bibinfo {author} {\bibfnamefont {H.-F.}\
  \bibnamefont {Wang}},\ }\bibfield  {title} {\bibinfo {title} {Enhanced photon
  blockade in an optomechanical system with parametric amplification},\ }\href
  {https://doi.org/10.1364/ol.392514} {\bibfield  {journal} {\bibinfo
  {journal} {Opt. Lett.}\ }\textbf {\bibinfo {volume} {45}},\ \bibinfo {pages}
  {2604} (\bibinfo {year} {2020})}\BibitemShut {NoStop}%
\bibitem [{\citenamefont {Durkin}\ \emph {et~al.}(2002)\citenamefont {Durkin},
  \citenamefont {Simon},\ and\ \citenamefont {Bouwmeester}}]{DSB02}%
  \BibitemOpen
  \bibfield  {author} {\bibinfo {author} {\bibfnamefont {G.~A.}\ \bibnamefont
  {Durkin}}, \bibinfo {author} {\bibfnamefont {C.}~\bibnamefont {Simon}},\ and\
  \bibinfo {author} {\bibfnamefont {D.}~\bibnamefont {Bouwmeester}},\
  }\bibfield  {title} {\bibinfo {title} {Multiphoton entanglement concentration
  and quantum cryptography},\ }\href
  {https://doi.org/10.1103/physrevlett.88.187902} {\bibfield  {journal}
  {\bibinfo  {journal} {Phys. Rev. Lett.}\ }\textbf {\bibinfo {volume} {88}},\
  \bibinfo {pages} {187902} (\bibinfo {year} {2002})}\BibitemShut {NoStop}%
\bibitem [{\citenamefont {Nagata}\ \emph {et~al.}(2007)\citenamefont {Nagata},
  \citenamefont {Okamoto}, \citenamefont {O{\textquotesingle}Brien},
  \citenamefont {Sasaki},\ and\ \citenamefont {Takeuchi}}]{NOO+07}%
  \BibitemOpen
  \bibfield  {author} {\bibinfo {author} {\bibfnamefont {T.}~\bibnamefont
  {Nagata}}, \bibinfo {author} {\bibfnamefont {R.}~\bibnamefont {Okamoto}},
  \bibinfo {author} {\bibfnamefont {J.~L.}\ \bibnamefont
  {O{\textquotesingle}Brien}}, \bibinfo {author} {\bibfnamefont
  {K.}~\bibnamefont {Sasaki}},\ and\ \bibinfo {author} {\bibfnamefont
  {S.}~\bibnamefont {Takeuchi}},\ }\bibfield  {title} {\bibinfo {title}
  {Beating the standard quantum limit with four-entangled photons},\ }\href
  {https://doi.org/10.1126/science.1138007} {\bibfield  {journal} {\bibinfo
  {journal} {Science}\ }\textbf {\bibinfo {volume} {316}},\ \bibinfo {pages}
  {726--729} (\bibinfo {year} {2007})}\BibitemShut {NoStop}%
\bibitem [{\citenamefont {Lombardi}\ \emph {et~al.}(2002)\citenamefont
  {Lombardi}, \citenamefont {Sciarrino}, \citenamefont {Popescu},\ and\
  \citenamefont {{De Martini}}}]{LSP+02}%
  \BibitemOpen
  \bibfield  {author} {\bibinfo {author} {\bibfnamefont {E.}~\bibnamefont
  {Lombardi}}, \bibinfo {author} {\bibfnamefont {F.}~\bibnamefont {Sciarrino}},
  \bibinfo {author} {\bibfnamefont {S.}~\bibnamefont {Popescu}},\ and\ \bibinfo
  {author} {\bibfnamefont {F.}~\bibnamefont {{De Martini}}},\ }\bibfield
  {title} {\bibinfo {title} {Teleportation of a vacuum{\textendash}one-photon
  qubit},\ }\href {https://doi.org/10.1103/physrevlett.88.070402} {\bibfield
  {journal} {\bibinfo  {journal} {Phys. Rev. Lett.}\ }\textbf {\bibinfo
  {volume} {88}},\ \bibinfo {pages} {070402} (\bibinfo {year}
  {2002})}\BibitemShut {NoStop}%
\bibitem [{\citenamefont {Ourjoumtsev}\ \emph {et~al.}(2007)\citenamefont
  {Ourjoumtsev}, \citenamefont {Jeong}, \citenamefont {Tualle-Brouri},\ and\
  \citenamefont {Grangier}}]{OJT+07}%
  \BibitemOpen
  \bibfield  {author} {\bibinfo {author} {\bibfnamefont {A.}~\bibnamefont
  {Ourjoumtsev}}, \bibinfo {author} {\bibfnamefont {H.}~\bibnamefont {Jeong}},
  \bibinfo {author} {\bibfnamefont {R.}~\bibnamefont {Tualle-Brouri}},\ and\
  \bibinfo {author} {\bibfnamefont {P.}~\bibnamefont {Grangier}},\ }\bibfield
  {title} {\bibinfo {title} {Generation of optical `schr\"{o}dinger cats' from
  photon number states},\ }\href {https://doi.org/10.1038/nature06054}
  {\bibfield  {journal} {\bibinfo  {journal} {Nature}\ }\textbf {\bibinfo
  {volume} {448}},\ \bibinfo {pages} {784--786} (\bibinfo {year}
  {2007})}\BibitemShut {NoStop}%
\bibitem [{\citenamefont {Dell'Anno}\ \emph {et~al.}(2006)\citenamefont
  {Dell'Anno}, \citenamefont {{De Siena}},\ and\ \citenamefont
  {Illuminati}}]{DDI06}%
  \BibitemOpen
  \bibfield  {author} {\bibinfo {author} {\bibfnamefont {F.}~\bibnamefont
  {Dell'Anno}}, \bibinfo {author} {\bibfnamefont {S.}~\bibnamefont {{De
  Siena}}},\ and\ \bibinfo {author} {\bibfnamefont {F.}~\bibnamefont
  {Illuminati}},\ }\bibfield  {title} {\bibinfo {title} {Multiphoton quantum
  optics and quantum state engineering},\ }\href
  {https://doi.org/10.1016/j.physrep.2006.01.004} {\bibfield  {journal}
  {\bibinfo  {journal} {Phys. Rep.}\ }\textbf {\bibinfo {volume} {428}},\
  \bibinfo {pages} {53--168} (\bibinfo {year} {2006})}\BibitemShut {NoStop}%
\bibitem [{\citenamefont {Montenegro}\ \emph {et~al.}(2019)\citenamefont
  {Montenegro}, \citenamefont {Ferraro},\ and\ \citenamefont {Bose}}]{MFB19}%
  \BibitemOpen
  \bibfield  {author} {\bibinfo {author} {\bibfnamefont {V.}~\bibnamefont
  {Montenegro}}, \bibinfo {author} {\bibfnamefont {A.}~\bibnamefont
  {Ferraro}},\ and\ \bibinfo {author} {\bibfnamefont {S.}~\bibnamefont
  {Bose}},\ }\bibfield  {title} {\bibinfo {title} {Enabling entanglement
  distillation via optomechanics},\ }\href
  {https://doi.org/10.1103/physreva.100.042310} {\bibfield  {journal} {\bibinfo
   {journal} {Phys. Rev. A}\ }\textbf {\bibinfo {volume} {100}},\ \bibinfo
  {pages} {042310} (\bibinfo {year} {2019})}\BibitemShut {NoStop}%
\bibitem [{\citenamefont {Kenfack}\ and\ \citenamefont {{\.
  Z}yczkowski}(2004)}]{KZ04}%
  \BibitemOpen
  \bibfield  {author} {\bibinfo {author} {\bibfnamefont {A.}~\bibnamefont
  {Kenfack}}\ and\ \bibinfo {author} {\bibfnamefont {K.}~\bibnamefont {{\.
  Z}yczkowski}},\ }\bibfield  {title} {\bibinfo {title} {Negativity of the
  wigner function as an indicator of non-classicality},\ }\href
  {https://doi.org/10.1088/1464-4266/6/10/003} {\bibfield  {journal} {\bibinfo
  {journal} {J. Opt. B: Quantum Semiclassical Opt.}\ }\textbf {\bibinfo
  {volume} {6}},\ \bibinfo {pages} {396--404} (\bibinfo {year}
  {2004})}\BibitemShut {NoStop}%
\bibitem [{\citenamefont {Leo{\'{n}}ski}(1996)}]{Leo96}%
  \BibitemOpen
  \bibfield  {author} {\bibinfo {author} {\bibfnamefont {W.}~\bibnamefont
  {Leo{\'{n}}ski}},\ }\bibfield  {title} {\bibinfo {title} {Fock states in a
  kerr medium with parametric pumping},\ }\href
  {https://doi.org/10.1103/physreva.54.3369} {\bibfield  {journal} {\bibinfo
  {journal} {Phys. Rev. A}\ }\textbf {\bibinfo {volume} {54}},\ \bibinfo
  {pages} {3369--3372} (\bibinfo {year} {1996})}\BibitemShut {NoStop}%
\bibitem [{\citenamefont {Rom{\'{a}}n-Ancheyta}\ \emph
  {et~al.}(2013)\citenamefont {Rom{\'{a}}n-Ancheyta}, \citenamefont
  {Guti{\'{e}}rrez},\ and\ \citenamefont {R{\'{e}}camier}}]{RGR13}%
  \BibitemOpen
  \bibfield  {author} {\bibinfo {author} {\bibfnamefont {R.}~\bibnamefont
  {Rom{\'{a}}n-Ancheyta}}, \bibinfo {author} {\bibfnamefont {C.~G.}\
  \bibnamefont {Guti{\'{e}}rrez}},\ and\ \bibinfo {author} {\bibfnamefont
  {J.}~\bibnamefont {R{\'{e}}camier}},\ }\bibfield  {title} {\bibinfo {title}
  {Photon-added nonlinear coherent states for a one-mode field in a kerr
  medium},\ }\href {https://doi.org/10.1364/josab.31.000038} {\bibfield
  {journal} {\bibinfo  {journal} {J. Opt. Soc. Am. B}\ }\textbf {\bibinfo
  {volume} {31}},\ \bibinfo {pages} {38} (\bibinfo {year} {2013})}\BibitemShut
  {NoStop}%
\bibitem [{\citenamefont {Ge}\ and\ \citenamefont {Zubairy}(2020)}]{GZ20}%
  \BibitemOpen
  \bibfield  {author} {\bibinfo {author} {\bibfnamefont {W.}~\bibnamefont
  {Ge}}\ and\ \bibinfo {author} {\bibfnamefont {M.~S.}\ \bibnamefont
  {Zubairy}},\ }\bibfield  {title} {\bibinfo {title} {Evaluating single-mode
  nonclassicality},\ }\href {https://doi.org/10.1103/physreva.102.043703}
  {\bibfield  {journal} {\bibinfo  {journal} {Phys. Rev. A}\ }\textbf {\bibinfo
  {volume} {102}},\ \bibinfo {pages} {043703} (\bibinfo {year}
  {2020})}\BibitemShut {NoStop}%
\bibitem [{\citenamefont {Mancini}\ \emph {et~al.}(1997)\citenamefont
  {Mancini}, \citenamefont {Man{\textquotesingle}ko},\ and\ \citenamefont
  {Tombesi}}]{MMT97}%
  \BibitemOpen
  \bibfield  {author} {\bibinfo {author} {\bibfnamefont {S.}~\bibnamefont
  {Mancini}}, \bibinfo {author} {\bibfnamefont {V.~I.}\ \bibnamefont
  {Man{\textquotesingle}ko}},\ and\ \bibinfo {author} {\bibfnamefont
  {P.}~\bibnamefont {Tombesi}},\ }\bibfield  {title} {\bibinfo {title}
  {Ponderomotive control of quantum macroscopic coherence},\ }\href
  {https://doi.org/10.1103/physreva.55.3042} {\bibfield  {journal} {\bibinfo
  {journal} {Phys. Rev. A}\ }\textbf {\bibinfo {volume} {55}},\ \bibinfo
  {pages} {3042--3050} (\bibinfo {year} {1997})}\BibitemShut {NoStop}%
\bibitem [{\citenamefont {Magnus}(1954)}]{Mag54}%
  \BibitemOpen
  \bibfield  {author} {\bibinfo {author} {\bibfnamefont {W.}~\bibnamefont
  {Magnus}},\ }\bibfield  {title} {\bibinfo {title} {On the exponential
  solution of differential equations for a linear operator},\ }\href
  {https://doi.org/10.1002/cpa.3160070404} {\bibfield  {journal} {\bibinfo
  {journal} {Commun. Pure Appl. Math}\ }\textbf {\bibinfo {volume} {7}},\
  \bibinfo {pages} {649--673} (\bibinfo {year} {1954})}\BibitemShut {NoStop}%
\bibitem [{\citenamefont {Johansson}\ \emph {et~al.}(2013)\citenamefont
  {Johansson}, \citenamefont {Nation},\ and\ \citenamefont {Nori}}]{Qutip1}%
  \BibitemOpen
  \bibfield  {author} {\bibinfo {author} {\bibfnamefont {J.}~\bibnamefont
  {Johansson}}, \bibinfo {author} {\bibfnamefont {P.}~\bibnamefont {Nation}},\
  and\ \bibinfo {author} {\bibfnamefont {F.}~\bibnamefont {Nori}},\ }\bibfield
  {title} {\bibinfo {title} {{QuTiP} 2: A python framework for the dynamics of
  open quantum systems},\ }\href {https://doi.org/10.1016/j.cpc.2012.11.019}
  {\bibfield  {journal} {\bibinfo  {journal} {Comput. Phys. Commun.}\ }\textbf
  {\bibinfo {volume} {184}},\ \bibinfo {pages} {1234--1240} (\bibinfo {year}
  {2013})}\BibitemShut {NoStop}%
\bibitem [{\citenamefont {Johansson}\ \emph {et~al.}(2012)\citenamefont
  {Johansson}, \citenamefont {Nation},\ and\ \citenamefont {Nori}}]{Qutip2}%
  \BibitemOpen
  \bibfield  {author} {\bibinfo {author} {\bibfnamefont {J.}~\bibnamefont
  {Johansson}}, \bibinfo {author} {\bibfnamefont {P.}~\bibnamefont {Nation}},\
  and\ \bibinfo {author} {\bibfnamefont {F.}~\bibnamefont {Nori}},\ }\bibfield
  {title} {\bibinfo {title} {{QuTiP}: An open-source python framework for the
  dynamics of open quantum systems},\ }\href
  {https://doi.org/10.1016/j.cpc.2012.02.021} {\bibfield  {journal} {\bibinfo
  {journal} {Comput. Phys. Commun.}\ }\textbf {\bibinfo {volume} {183}},\
  \bibinfo {pages} {1760--1772} (\bibinfo {year} {2012})}\BibitemShut {NoStop}%
\bibitem [{Note1()}]{Note1}%
  \BibitemOpen
  \bibinfo {note} {Simulations of dissipative dynamics are performed bases on
  the Trotter-Suzuki decomposition with terms induced by time-independent
  effective Hamiltonian and terms induced by Lindbladians for the dissipative
  part of the dynamics.}\BibitemShut {Stop}%
\bibitem [{\citenamefont {Dezfouli}\ \emph {et~al.}(2019)\citenamefont
  {Dezfouli}, \citenamefont {Gordon},\ and\ \citenamefont {Hughes}}]{DGH19}%
  \BibitemOpen
  \bibfield  {author} {\bibinfo {author} {\bibfnamefont {M.~K.}\ \bibnamefont
  {Dezfouli}}, \bibinfo {author} {\bibfnamefont {R.}~\bibnamefont {Gordon}},\
  and\ \bibinfo {author} {\bibfnamefont {S.}~\bibnamefont {Hughes}},\
  }\bibfield  {title} {\bibinfo {title} {Molecular optomechanics in the
  anharmonic cavity-{QED} regime using hybrid metal{\textendash}dielectric
  cavity modes},\ }\href {https://doi.org/10.1021/acsphotonics.8b01091}
  {\bibfield  {journal} {\bibinfo  {journal} {ACS Photonics}\ }\textbf
  {\bibinfo {volume} {6}},\ \bibinfo {pages} {1400--1408} (\bibinfo {year}
  {2019})}\BibitemShut {NoStop}%
\bibitem [{\citenamefont {Hu}\ \emph {et~al.}(2015)\citenamefont {Hu},
  \citenamefont {Huang}, \citenamefont {Liao}, \citenamefont {Tian},\ and\
  \citenamefont {Goan}}]{HHL+15}%
  \BibitemOpen
  \bibfield  {author} {\bibinfo {author} {\bibfnamefont {D.}~\bibnamefont
  {Hu}}, \bibinfo {author} {\bibfnamefont {S.-Y.}\ \bibnamefont {Huang}},
  \bibinfo {author} {\bibfnamefont {J.-Q.}\ \bibnamefont {Liao}}, \bibinfo
  {author} {\bibfnamefont {L.}~\bibnamefont {Tian}},\ and\ \bibinfo {author}
  {\bibfnamefont {H.-S.}\ \bibnamefont {Goan}},\ }\bibfield  {title} {\bibinfo
  {title} {Quantum coherence in ultrastrong optomechanics},\ }\href
  {https://doi.org/10.1103/physreva.91.013812} {\bibfield  {journal} {\bibinfo
  {journal} {Phys. Rev. A}\ }\textbf {\bibinfo {volume} {91}},\ \bibinfo
  {pages} {013812} (\bibinfo {year} {2015})}\BibitemShut {NoStop}%
\bibitem [{\citenamefont {Kronwald}\ \emph {et~al.}(2013)\citenamefont
  {Kronwald}, \citenamefont {Marquardt},\ and\ \citenamefont {Clerk}}]{KMC13}%
  \BibitemOpen
  \bibfield  {author} {\bibinfo {author} {\bibfnamefont {A.}~\bibnamefont
  {Kronwald}}, \bibinfo {author} {\bibfnamefont {F.}~\bibnamefont
  {Marquardt}},\ and\ \bibinfo {author} {\bibfnamefont {A.~A.}\ \bibnamefont
  {Clerk}},\ }\bibfield  {title} {\bibinfo {title} {Arbitrarily large
  steady-state bosonic squeezing via dissipation},\ }\href
  {https://doi.org/10.1103/physreva.88.063833} {\bibfield  {journal} {\bibinfo
  {journal} {Phys. Rev. A}\ }\textbf {\bibinfo {volume} {88}},\ \bibinfo
  {pages} {063833} (\bibinfo {year} {2013})}\BibitemShut {NoStop}%
\bibitem [{\citenamefont {Arenz}\ \emph {et~al.}(2013)\citenamefont {Arenz},
  \citenamefont {Cormick}, \citenamefont {Vitali},\ and\ \citenamefont
  {Morigi}}]{ACV+13}%
  \BibitemOpen
  \bibfield  {author} {\bibinfo {author} {\bibfnamefont {C.}~\bibnamefont
  {Arenz}}, \bibinfo {author} {\bibfnamefont {C.}~\bibnamefont {Cormick}},
  \bibinfo {author} {\bibfnamefont {D.}~\bibnamefont {Vitali}},\ and\ \bibinfo
  {author} {\bibfnamefont {G.}~\bibnamefont {Morigi}},\ }\bibfield  {title}
  {\bibinfo {title} {Generation of two-mode entangled states by quantum
  reservoir engineering},\ }\href
  {https://doi.org/10.1088/0953-4075/46/22/224001} {\bibfield  {journal}
  {\bibinfo  {journal} {J. Phys. B: At., Mol. Opt. Phys.}\ }\textbf {\bibinfo
  {volume} {46}},\ \bibinfo {pages} {224001} (\bibinfo {year}
  {2013})}\BibitemShut {NoStop}%
\bibitem [{\citenamefont {Asjad}\ and\ \citenamefont {Vitali}(2014)}]{AV14}%
  \BibitemOpen
  \bibfield  {author} {\bibinfo {author} {\bibfnamefont {M.}~\bibnamefont
  {Asjad}}\ and\ \bibinfo {author} {\bibfnamefont {D.}~\bibnamefont {Vitali}},\
  }\bibfield  {title} {\bibinfo {title} {Reservoir engineering of a mechanical
  resonator: generating a macroscopic superposition state and monitoring its
  decoherence},\ }\href {https://doi.org/10.1088/0953-4075/47/4/045502}
  {\bibfield  {journal} {\bibinfo  {journal} {J. Phys. B: At., Mol. Opt.
  Phys.}\ }\textbf {\bibinfo {volume} {47}},\ \bibinfo {pages} {045502}
  (\bibinfo {year} {2014})}\BibitemShut {NoStop}%
\bibitem [{\citenamefont {Clerk}\ \emph {et~al.}(2010)\citenamefont {Clerk},
  \citenamefont {Devoret}, \citenamefont {Girvin}, \citenamefont {Marquardt},\
  and\ \citenamefont {Schoelkopf}}]{CDG+10}%
  \BibitemOpen
  \bibfield  {author} {\bibinfo {author} {\bibfnamefont {A.~A.}\ \bibnamefont
  {Clerk}}, \bibinfo {author} {\bibfnamefont {M.~H.}\ \bibnamefont {Devoret}},
  \bibinfo {author} {\bibfnamefont {S.~M.}\ \bibnamefont {Girvin}}, \bibinfo
  {author} {\bibfnamefont {F.}~\bibnamefont {Marquardt}},\ and\ \bibinfo
  {author} {\bibfnamefont {R.~J.}\ \bibnamefont {Schoelkopf}},\ }\bibfield
  {title} {\bibinfo {title} {Introduction to quantum noise, measurement, and
  amplification},\ }\href {https://doi.org/10.1103/revmodphys.82.1155}
  {\bibfield  {journal} {\bibinfo  {journal} {Rev. Mod. Phys.}\ }\textbf
  {\bibinfo {volume} {82}},\ \bibinfo {pages} {1155--1208} (\bibinfo {year}
  {2010})}\BibitemShut {NoStop}%
\bibitem [{\citenamefont {Harvey}(2017)}]{imperialHPC}%
  \BibitemOpen
  \bibfield  {author} {\bibinfo {author} {\bibfnamefont {M.}~\bibnamefont
  {Harvey}},\ }\href {https://doi.org/10.14469/HPC/2232} {\bibinfo {title}
  {Imperial college research computing service}} (\bibinfo {year}
  {2017})\BibitemShut {NoStop}%
\end{thebibliography}%

\end{document}